\documentclass[aps,tightenlines,superscriptaddress,twocolumn]{revtex4-1}

\usepackage{graphicx}
\usepackage{amssymb}
\usepackage{multirow}
\usepackage{float}
\usepackage{xcolor}

\usepackage{lineno}

\newcommand{\sNN}{\sqrt{s_{_{\rm NN}}}}
\begin{document}


\title{Elliptic Flow of Heavy-Flavor Decay Electrons in Au+Au Collisions at $\sqrt{s_{_{\rm NN}}}$ = 27 and 54.4 GeV at RHIC}

\affiliation{Abilene Christian University, Abilene, Texas   79699}
\affiliation{AGH University of Science and Technology, FPACS, Cracow 30-059, Poland}
\affiliation{Argonne National Laboratory, Argonne, Illinois 60439}
\affiliation{American University in Cairo, New Cairo 11835, Egypt}
\affiliation{Ball State University, Muncie, Indiana, 47306}
\affiliation{Brookhaven National Laboratory, Upton, New York 11973}
\affiliation{University of Calabria \& INFN-Cosenza, Rende 87036, Italy}
\affiliation{University of California, Berkeley, California 94720}
\affiliation{University of California, Davis, California 95616}
\affiliation{University of California, Los Angeles, California 90095}
\affiliation{University of California, Riverside, California 92521}
\affiliation{Central China Normal University, Wuhan, Hubei 430079 }
\affiliation{University of Illinois at Chicago, Chicago, Illinois 60607}
\affiliation{Creighton University, Omaha, Nebraska 68178}
\affiliation{Czech Technical University in Prague, FNSPE, Prague 115 19, Czech Republic}
\affiliation{Technische Universit\"at Darmstadt, Darmstadt 64289, Germany}
\affiliation{National Institute of Technology Durgapur, Durgapur - 713209, India}
\affiliation{ELTE E\"otv\"os Lor\'and University, Budapest, Hungary H-1117}
\affiliation{Frankfurt Institute for Advanced Studies FIAS, Frankfurt 60438, Germany}
\affiliation{Fudan University, Shanghai, 200433 }
\affiliation{University of Heidelberg, Heidelberg 69120, Germany }
\affiliation{University of Houston, Houston, Texas 77204}
\affiliation{Huzhou University, Huzhou, Zhejiang  313000}
\affiliation{Indian Institute of Science Education and Research (IISER), Berhampur 760010 , India}
\affiliation{Indian Institute of Science Education and Research (IISER) Tirupati, Tirupati 517507, India}
\affiliation{Indian Institute Technology, Patna, Bihar 801106, India}
\affiliation{Indiana University, Bloomington, Indiana 47408}
\affiliation{Institute of Modern Physics, Chinese Academy of Sciences, Lanzhou, Gansu 730000 }
\affiliation{University of Jammu, Jammu 180001, India}
\affiliation{Kent State University, Kent, Ohio 44242}
\affiliation{University of Kentucky, Lexington, Kentucky 40506-0055}
\affiliation{Lawrence Berkeley National Laboratory, Berkeley, California 94720}
\affiliation{Lehigh University, Bethlehem, Pennsylvania 18015}
\affiliation{Max-Planck-Institut f\"ur Physik, Munich 80805, Germany}
\affiliation{Michigan State University, East Lansing, Michigan 48824}
\affiliation{National Institute of Science Education and Research, HBNI, Jatni 752050, India}
\affiliation{National Cheng Kung University, Tainan 70101 }
\affiliation{Nuclear Physics Institute of the CAS, Rez 250 68, Czech Republic}
\affiliation{The Ohio State University, Columbus, Ohio 43210}
\affiliation{Institute of Nuclear Physics PAN, Cracow 31-342, Poland}
\affiliation{Panjab University, Chandigarh 160014, India}
\affiliation{Purdue University, West Lafayette, Indiana 47907}
\affiliation{Rice University, Houston, Texas 77251}
\affiliation{Rutgers University, Piscataway, New Jersey 08854}
\affiliation{Universidade de S\~ao Paulo, S\~ao Paulo, Brazil 05314-970}
\affiliation{University of Science and Technology of China, Hefei, Anhui 230026}
\affiliation{South China Normal University, Guangzhou, Guangdong 510631}
\affiliation{Sejong University, Seoul, 05006, South Korea}
\affiliation{Shandong University, Qingdao, Shandong 266237}
\affiliation{Shanghai Institute of Applied Physics, Chinese Academy of Sciences, Shanghai 201800}
\affiliation{Southern Connecticut State University, New Haven, Connecticut 06515}
\affiliation{State University of New York, Stony Brook, New York 11794}
\affiliation{Instituto de Alta Investigaci\'on, Universidad de Tarapac\'a, Arica 1000000, Chile}
\affiliation{Temple University, Philadelphia, Pennsylvania 19122}
\affiliation{Texas A\&M University, College Station, Texas 77843}
\affiliation{University of Texas, Austin, Texas 78712}
\affiliation{Tsinghua University, Beijing 100084}
\affiliation{University of Tsukuba, Tsukuba, Ibaraki 305-8571, Japan}
\affiliation{University of Chinese Academy of Sciences, Beijing, 101408}
\affiliation{United States Naval Academy, Annapolis, Maryland 21402}
\affiliation{Valparaiso University, Valparaiso, Indiana 46383}
\affiliation{Variable Energy Cyclotron Centre, Kolkata 700064, India}
\affiliation{Warsaw University of Technology, Warsaw 00-661, Poland}
\affiliation{Wayne State University, Detroit, Michigan 48201}
\affiliation{Yale University, New Haven, Connecticut 06520}

\author{M.~I.~Abdulhamid}\affiliation{American University in Cairo, New Cairo 11835, Egypt}
\author{B.~E.~Aboona}\affiliation{Texas A\&M University, College Station, Texas 77843}
\author{J.~Adam}\affiliation{Czech Technical University in Prague, FNSPE, Prague 115 19, Czech Republic}
\author{L.~Adamczyk}\affiliation{AGH University of Science and Technology, FPACS, Cracow 30-059, Poland}
\author{J.~R.~Adams}\affiliation{The Ohio State University, Columbus, Ohio 43210}
\author{I.~Aggarwal}\affiliation{Panjab University, Chandigarh 160014, India}
\author{M.~M.~Aggarwal}\affiliation{Panjab University, Chandigarh 160014, India}
\author{Z.~Ahammed}\affiliation{Variable Energy Cyclotron Centre, Kolkata 700064, India}
\author{D.~M.~Anderson}\affiliation{Texas A\&M University, College Station, Texas 77843}
\author{E.~C.~Aschenauer}\affiliation{Brookhaven National Laboratory, Upton, New York 11973}
% ...

\collaboration{STAR Collaboration}\noaffiliation

\date{\today}
\begin{abstract}
We report on new measurements of elliptic flow ($v_2$) of electrons from heavy-flavor hadron decays at mid-rapidity ($|y|<0.8$) in Au+Au collisions at $\sqrt{s_{_{\rm NN}}}$ = 27 and 54.4\,GeV from the STAR experiment.
Heavy-flavor decay electrons ($e^{\rm HF}$) in Au+Au collisions at $\sqrt{s_{_{\rm NN}}}$ = 54.4\,GeV exhibit a non-zero $v_2$ in the transverse momentum ($p_{\rm T}$) region of $p_{\rm T}<$ 2\,GeV/$c$ with the magnitude comparable to that at $\sNN=200$ \,GeV. 
The measured $e^{\rm HF}$ $v_2$ at 54.4\,GeV is also consistent with the expectation of their parent charm hadron $v_2$ following number-of-constituent-quark scaling as other light and strange flavor hadrons at this energy. 
These suggest that charm quarks gain significant collectivity through the evolution of the QCD medium and may reach local thermal equilibrium in Au+Au collisions at $\sNN=54.4$\,GeV.
The measured $e^{\rm HF}$ $v_2$ in Au+Au collisions at $\sNN=$ 27\,GeV is consistent with zero within large uncertainties. 
The energy dependence of $v_2$ for different flavor particles ($\pi,\phi,D^{0}/e^{\rm HF}$) shows an indication of quark mass hierarchy in reaching thermalization in high-energy nuclear collisions.
\end{abstract}



\maketitle

\section{Introduction}
\label{sec:intro}
Heavy-ion collisions offer a unique environment to study quantum chromodynamics (QCD) in a laboratory, particularly at extremely high temperature and density conditions.
Experiments at the Relativistic Heavy Ion Collider (RHIC) and Large Hadron Collider (LHC) have demonstrated that a novel QCD matter, namely the Quark-Gluon Plasma (QGP), is created in ultra-relativistic heavy-ion collisions~\cite{Adams:2005dq,Adcox:2004mh,Muller:2012zq}. One critical mission of the current RHIC and LHC heavy-ion experiments is to determine the microscopic properties of the QGP medium quantitatively. Heavy-flavor quarks ($c$, $b$) have unique roles in this direction primarily due to their large mass.

Heavy-flavor quarks are predominantly produced through initial hard scattering processes in heavy-ion collisions. Their thermal relaxation time is expected to be comparable to or longer than the typical lifetime of the QGP medium created at the RHIC and LHC~\cite{Svetitsky:1987gq,Moore:2004tg,Rapp:2009my}. 
The collectivity of heavy-flavor quarks, especially in the low transverse momentum ($p_{\rm T}$) region, is sensitive to the strongly coupled QGP medium transport parameter, called the heavy-flavor quark spatial diffusion coefficient (${\cal{D}}_s$)~\cite{Akiba:2015jwa}.

In heavy-ion collisions, particle collectivity is often characterized by anisotropic parameters $v_n$, the $n$-th harmonic coefficient in the Fourier decomposition of the particles azimuthal distribution ($dN/d\phi$) with respect to the event planes $\Psi_{n}$~\cite{Poskanzer:1998yz,Voloshin:1994mz}:
\begin{equation}
   \frac{dN}{d\phi}\varpropto1+2\sum^{\infty}_{n=1}v_{n}\cos[n(\phi-\Psi_{n})].
\end{equation}
The second harmonic coefficient, $v_2$, is called elliptic flow. 

The charmed hadron elliptic flow~\cite{Adamczyk:2017xur,Acharya:2017qps,Sirunyan:2017plt} and the nuclear modification factor ($R_{\rm AA}$)~\cite{Adamczyk:2014uip,Adam:2018inb,Acharya:2018hre,Sirunyan:2017xss,ALICE:2021rxa} have been measured several times at top RHIC and LHC energies. 
Results show that charm hadron production is significantly suppressed  at high pT region and charm hadrons exhibit significant collectivity, indicating charm quarks are strongly coupled with the QGP medium.
Measurements using single leptons from heavy-flavor hadron decays at these energies provide similar observations~\cite{starHFe,Adare:2006nq,Acharya:2019mom,ATLAS:2020yxw}. Recent phenomenological models constrained by these results suggest that the dimensionless charm quark spatial diffusion coefficient $2\pi T{\cal{D}}_s$ is about 2--5 in the vicinity of the critical temperature while its temperature ($T$) dependence remains uncertain~\cite{Rapp:2018qla,Cao:2018ews,Dong:2019byy}. This value is consistent with quenched lattice QCD calculations within large uncertainties~\cite{Banerjee:2011ra,Ding:2012sp,Brambilla:2020siz}. The next important task of the heavy-flavor program is to further constrain the diffusion coefficient and investigate its dependence on momentum, temperature, as well as baryon chemical potential ($\mu_B$). Measuring heavy-flavor quark collectivity below the RHIC top energy offers new insights into the $T$ and $\mu_B$ dependence of the QGP transport parameter, ${\cal{D}}_s$.

While previous measurements exist from RHIC experiments on heavy-flavor decay electron $v_2$ in Au+Au collisions at $\sqrt{s_{\rm NN}} = 62.4$ and 39\,GeV~\cite{starHFe,phenix62HFe}, the accompanying large statistical and systematic uncertainties prevent firm conclusions on charm quark collectivity at energies below 200 GeV.
In this paper, we report new measurements of heavy-flavor decay electrons $v_2$ from Au+Au collisions at $\sqrt{s_{\rm NN}}$ = 54.4 and $27\, \rm GeV$ from the STAR experiment.
\section{Experimental Setup and Analysis Method}
\label{sec:exp}
The data utilized in this analysis is from Au+Au collisions at $\sqrt{s_{_{\rm NN}}}=$ 54.4 and 27 GeV collected by the STAR experiment in 2017 and 2018, respectively. For the $\sNN=54.4$\,GeV data, a minimum-bias trigger was used which was defined as the coincidence of the two zero-degree calorimeters (ZDC, $|\eta|>6.0$)~\cite{Judd:2018zbg,Adler:2000bd}, or the two vertex position detectors (VPD, $4.2<|\eta|<5.1$)~\cite{Judd:2018zbg,Llope:2014nva}. For the $\sNN=27$ GeV data, the minimum-bias triggered events also include those with the coincidence of the beam-beam counters (BBC, $2.2<|\eta|<5.0$) and having multiplicity recorded by the Time-of-Flight (TOF, $|\eta|<0.9$)~\cite{BONNER2003181} above a certain threshold~\cite{Judd:2018zbg}. The offline reconstructed collision vertex of each event is required to be within $\pm 35\, \rm cm$ of the nominal center of the STAR detector along the beam direction. The centrality is determined by comparing charged particle multiplicity in $|\eta|<0.5$ with a Monte Carlo Glauber model simulation \cite{Miller:2007ri,STAR:2009sxc}. For this analysis, a centrality range of 0-60\% is selected to utilize statistics fully. There are 5.7$\times 10^8$ and 2.4$\times 10^8$ events passing the selection mentioned above for the analysis at $\sNN=$ 54.4 and 27 GeV, respectively. The statistics of these data samples are more than a factor of 10 times larger compared to the data used in the previous STAR measurements of single electron $v_2$ at $\sNN=$ 62.4 and $39\,\rm GeV$, respectively \cite{starHFe}.

The Time Projection Chamber (TPC)~\cite{Anderson:2003ur} and the Time-of-Flight~\cite{Llope:2003ti} are the two main sub-detector systems used for tracking and particle identification.
Tracks are required to be reconstructed with at least 20 TPC hit points out of a maximum of 45. The ratio of the number of track hit points used for track reconstruction to the maximum possible hits must also be at least 52\% to reject split tracks. The distance-of-closest approach (DCA) of the tracks to the primary vertex of the tracks is required to be less than 1.5 cm to reduce the secondary electrons from photons converted in the detector material. Tracks are selected within pseudorapidity ranges $|\eta|<0.8$, azimuthal angle region of $-1.25<\phi<1.25$, and $1.95<|\phi|<\pi$ to suppress the electrons from photon conversion in the support structures of the Silicon Vertex Tracker (SVT) \cite{STAR:2002bzu} and the beam pipe. If not specified in the paper, the selection criteria used in the analysis, e.g. selection of electron tracks, photonic electron tagging, and event plane reconstruction, are the same for both collision energies.

\begin{figure}[h]
\centering\includegraphics[width=0.75\linewidth]{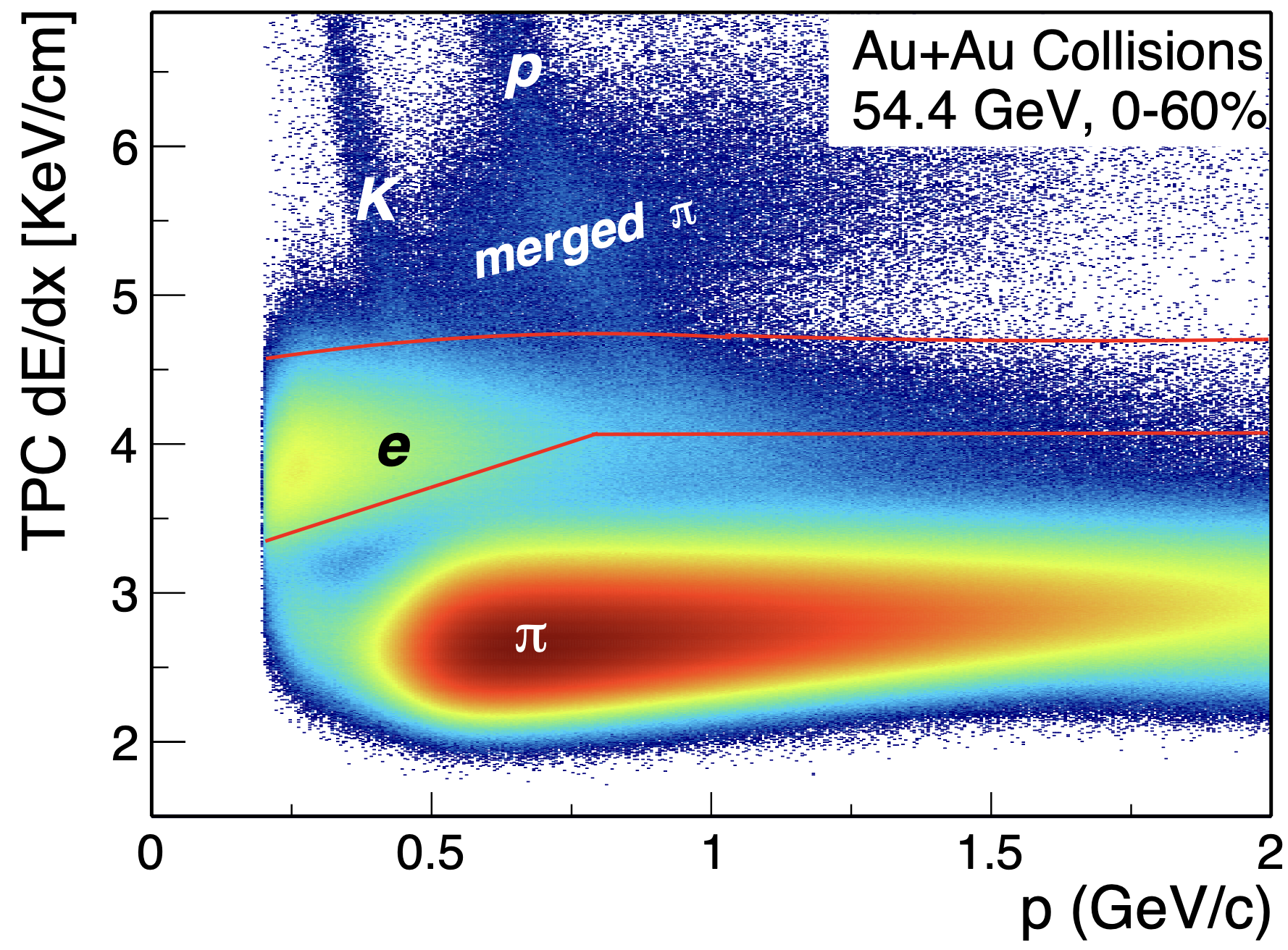}
\caption{The $dE/dx$ distribution of tracks as a function of momentum in Au+Au collisions at $\sNN=54.4$ GeV, after passing TOF electron selection criteria. The electron samples are selected within the two red lines.}
\label{fig:dEdx}
\end{figure}
\begin{figure}[h]
\centering\includegraphics[width=0.85\linewidth]{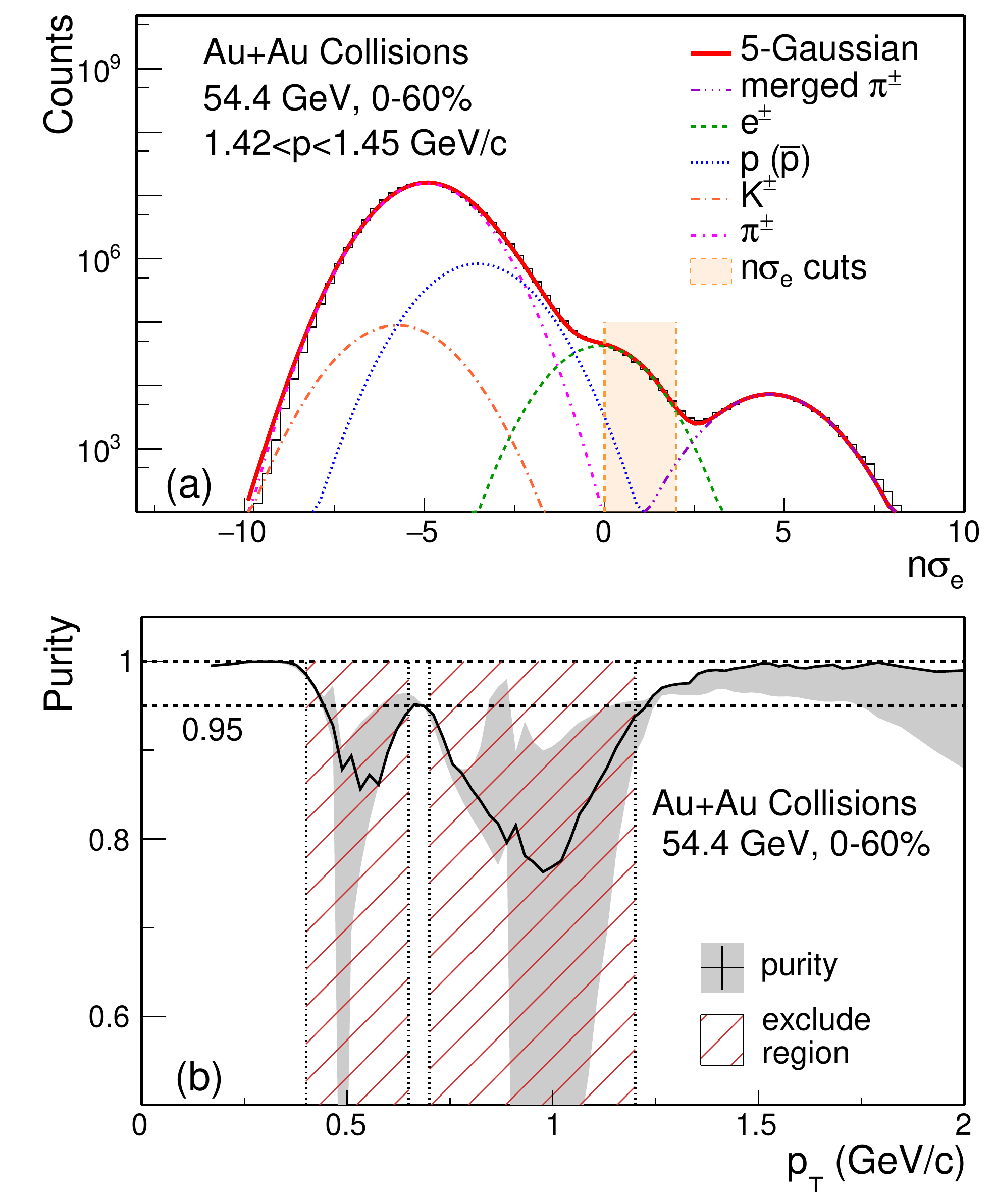}
\caption{(a) An example $n\sigma_{e}$ distribution with five-Gaussian fit (red solid curves) at $1.42<p<1.45$ GeV/c in Au+Au collisions at $\sNN=$ $54.4\, \rm GeV$. Contributions from different particle species are indicated as dashed or dot-dashed lines. The electron samples within the $n\sigma_{e}$ selection criteria are designated by the orange-filled area. (b) The purity of the inclusive electron candidates after both $dE/dx$ and TOF PID in Au+Au collisions at $\sqrt{s_{_{\rm NN}}}=$ $54.4\,\rm GeV$. The gray band represents systematic uncertainties.}
\label{fig:purity}
\end{figure}

\begin{figure*}[thbp]
\centering\includegraphics[width=0.8\linewidth]{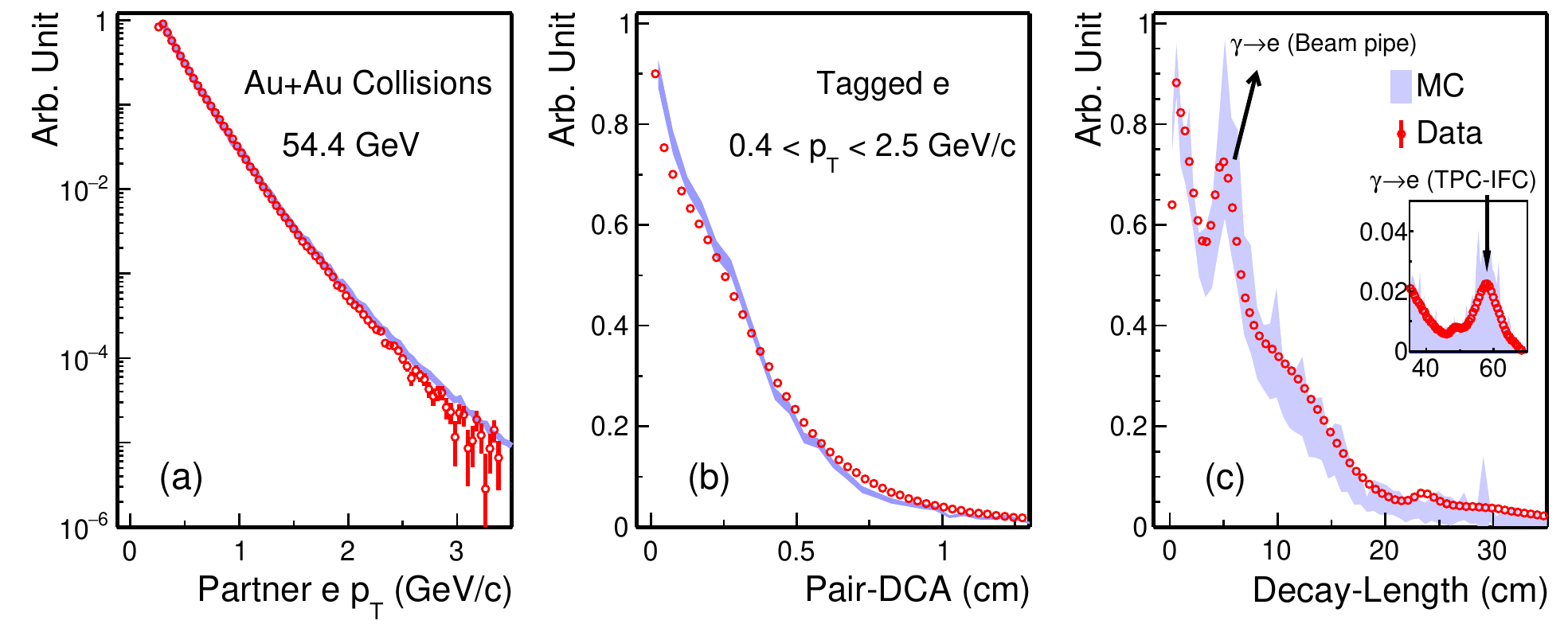}
\caption{Comparisons of $p_{\rm T}$ and topological distributions between data (open circles) and Monte Carlo (blue bands) at tagged electrons $0.4<p_{\rm T}<2.5$\,GeV/$c$ in Au+Au collisions at $\sqrt{s_{_{\rm NN}}}$ = 54.4\,GeV. (a) Photonic electron partner $p_{\rm T}$. (b) Electron pair DCA. (c) Position distance to primary vertex distributions. Peaks around 5 and 60 cm in panel (c) arise from photon conversion in the beam pipe and TPC inner field cage, respectively.}
\label{fig:dataMC}
\end{figure*}

In the following part of this section, we first describe how to identify electrons in our experiment and its purity correction. The electron candidates contain signals (heavy-flavor decay electrons, $e^{\rm HF}$) and various background sources that include electrons from photons converted in detector material and $\pi^0,\eta$ decays (photonic electrons), from vector meson decays and kaon weak decays. We describe in detail how to remove these background and correct for their contamination in the final elliptic flow measurement.

Electron tracks are identified using the inverse velocity ($1/\beta$) calculated from the path length and time of flight between the collision vertex point and the TOF detector and are required to satisfy $|1-1/\beta|<0.025$. Then electron candidate tracks are further selected by the ionization energy loss ($dE/dx$) \cite{BICHSEL2006154} in the TPC.
The $dE/dx$ distribution of the tracks that have passed $1/\beta$ cuts is shown in Fig. \ref{fig:dEdx}. The electron tracks are selected as $(p\times3.5-2.8)<n\sigma_{e}<2$ at $p<0.8$ GeV/$c$ and $0<n\sigma_{e}<2$ at $p>0.8$ GeV/$c$ where $n\sigma_{e}$ is the normalized $dE/dx$ \cite{XU201028}. $n\sigma_{e}$ is defined as $n\sigma_e = \ln[(dE/dx)^{meas}/(dE/dx)^{exp}]/R$, where $(dE/dx)^{meas}$ and $(dE/dx)^{exp}$ is the measured and theoretically expected $dE/dx$, respectively, and $R$ is the TPC resolution of $\ln[(dE/dx)^{meas}/(dE/dx)^{exp}]$ \cite{XU201028}. The candidates that pass all track quality and particle identification (PID) requirements are categorized as inclusive electron candidates. Both electrons and positrons are used in the analysis.

As indicated in Fig.\,\ref{fig:dEdx}, hadrons, including kaon, pion, proton, and the ``merged pions", contaminate our inclusive electron candidates. Merged pions are two pion tracks that cannot be separated due to the finite spatial resolution of the TPC. To evaluate hadron contamination, the $n\sigma_{e}$ distributions of pure hadron and electron samples are used as templates and described by Gaussian functions \cite{starHFe}. Then, the mean and width of the $n\sigma_{e}$ distribution of each particle species can be obtained from the Gaussian fitting to the above templates.
A multi-Gaussian function with fixed mean and width, and free amplitude for each component is used to fit the $n\sigma_{e}$ distribution of electron candidates that pass $1/\beta$ cuts. The fitting is done within narrow momentum intervals to ensure $n\sigma_{e}$ distributions of various particle species are close to being Gaussian distributed. Figure \ref{fig:purity}(a) shows an example of a multi-Gaussian fit at $1.42<p<1.45$ GeV/c for the $\sNN=$ $54.4\,\rm GeV$ analysis. The purity of inclusive electron candidates is calculated as the ratio of the electron yield over the yield of all candidates within the $n\sigma_{e}$ cuts used in the analysis. Electron purity is first evaluated as a function of momentum, and then transformed to the $p_{\rm T}$ dependence based on the correlation between inclusive electron $p_{\rm T}$ and its momentum. As shown in Fig. \ref{fig:dEdx}, the $dE/dx$ bands for kaon and proton cross with the electron band in certain momentum ranges ($p\sim 0.5$ GeV/$c$ for kaon and $p\sim 1$ GeV/$c$ for proton) resulting in significant drops of the electron purity, as seen in Fig.~\ref{fig:purity}(b).
The following sources of variance are included in estimating systematic uncertainty: (1) the changing of constraints on particle yields for the multi-Gaussian fitting; (2) the conditional pion selection from either $K_{S}^{0}\rightarrow\pi^+\pi^-$ or from TOF identification; (3) the alternation of the functions used to describe the pion $n\sigma_{e}$ distribution.
The estimated electron purity as a function of $p_{\rm T}$ is shown in Fig. \ref{fig:purity}(b).  We exclude the $p_{\rm T}$ ranges of $0.4<p_{\rm T}<0.65\,{\rm GeV}/c$ and $0.7<p_{\rm T}<1.2\,{\rm GeV}/c$ in $\sNN=$ $54.4\,\rm GeV$ measurements, and $0.4<p_{\rm T}<0.6\, {\rm GeV}/c$ and $0.7<p_{\rm T}<1.2\,{\rm GeV}/c$ in $\sNN=$ $27\,\rm GeV$ measurements. Since the electron $dE/dx$ band crosses with those for kaon and proton respectively in those $p_{\rm T}$ ranges and systematic uncertainties would otherwise greatly conceal results.

\begin{figure}[tb]
\centering\includegraphics[width=0.9\linewidth]{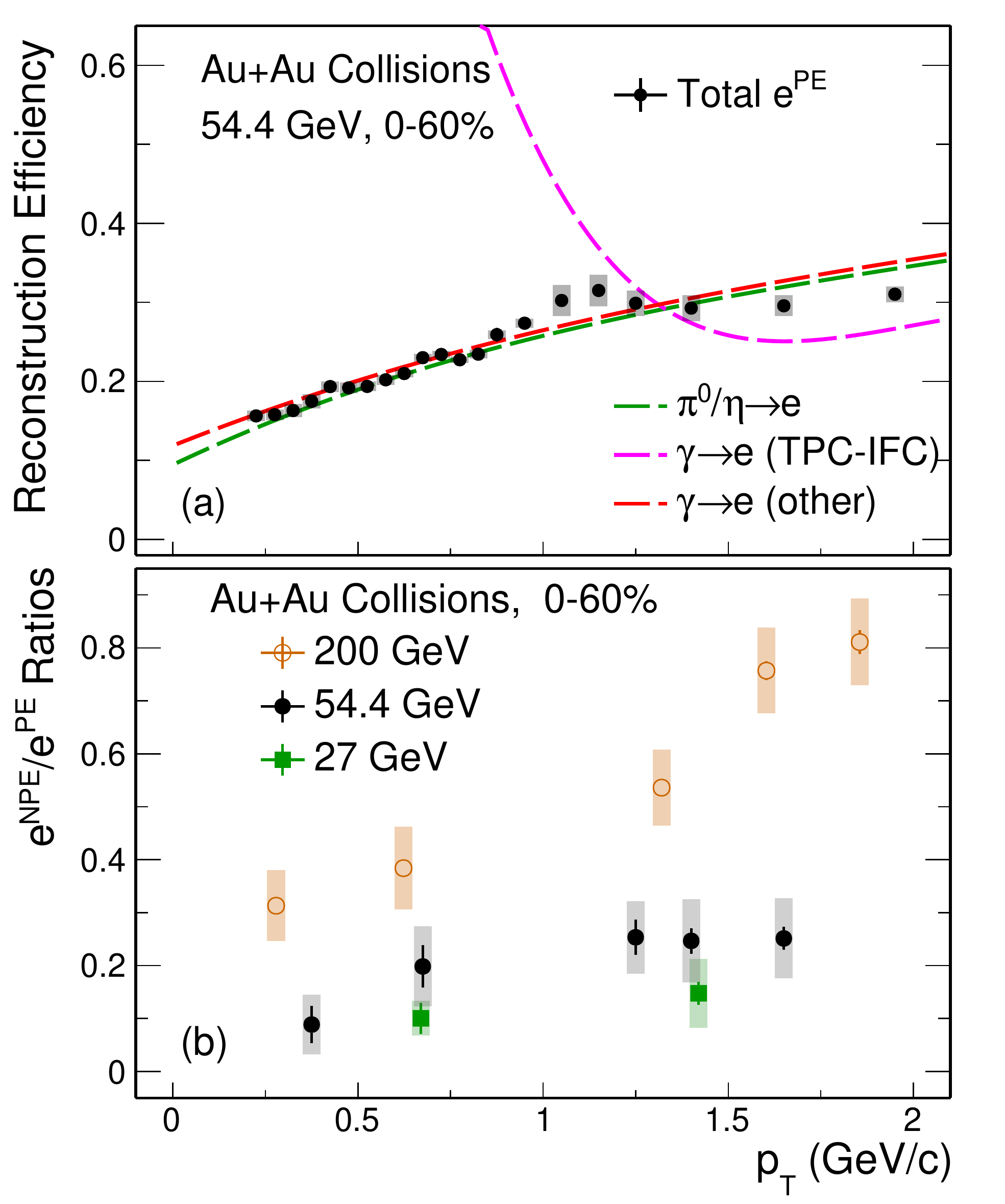}
\caption{(a) The total photonic electron reconstruction efficiency is shown as the solid points. Dashed lines depict the reconstruction efficiency of photonic electrons from various sources, including Dalitz decay electrons from $\pi^0$ and $\eta$ (green), photon conversion electrons that are converted in the TPC-IFC (magenta), conversions in other detector materials (red). (b) Non-photonic electrons ($e^{\rm NPE}$) to photonic electrons $e^{\rm PE}$ yield ratio as a function of tagged electron $p_{\rm T}$ in 0-60\% Au+Au collisions at $\sqrt{s_{_{\rm NN}}}$ = 200 (open circle) \cite{starHFe}, 54.4 (full circle), and 27 (full square) GeV. The data points at $\sNN=$ 200 GeV collisions \cite{starHFe} have excluded $\sim8\%$ contributions from $K_{e3}$. Boxes on data points depict systematic uncertainties. Data points from 27 GeV are shifted horizontally for clarity. The vertical bars and boxes denote the statistical and systematic uncertainties,  respectively.}
\label{fig:Eff}
\end{figure}

The dominant sources of background for heavy-flavor decay electrons are photonic electrons ($e^{\rm PE}$) from Dalitz decays of light mesons (predominantly $\pi^0$, $\eta$) and photon conversion in the detector material. The yield of non-photonic electrons (NPE) can be calculated as:
\begin{equation}
N^{\rm NPE} = \kappa\times N^{\rm inc}-N^{\rm PE},
\label{eq:n_NPE}
\end{equation}
where $\kappa$ is the electron purity. $N^{\rm inc}$ and $N^{\rm PE}$ are the yield of inclusive electrons and photonic electrons, respectively. The yield of photonic electrons ($N^{\rm PE}$) is evaluated by the following reconstruction method described in \cite{starHFe, Agakishiev:2011mr}. Inclusive electron tracks (called tagged electrons), are paired with opposite-sign partner electrons (Unlike-Sign) randomly in the same event. A tagged electron is regarded as the photonic electron candidate if the dielectron pair passes reconstruction cuts, which requires a pair DCA of less than 1 cm and a reconstructed invariant mass of less than 0.1\,GeV/$c^2$. Photonic electrons that are successfully tagged by dielectron reconstruction are called reconstructed photonic electrons ($e^{\rm reco}$). The combinatorial background is estimated by pairing tagged electrons with same-sign electrons (Like-Sign). The photonic electron yield is calculated statistically as follows:
\begin{equation}
N^{\rm PE}=(N^{\rm UL}-N^{\rm LS})/\varepsilon^{\rm reco}, 
\label{eq:n_pho}
\end{equation}
where $N^{\rm UL}$ and $N^{\rm LS}$ are the number of Unlike-Sign and Like-Sign electron pairs that have passed reconstruction cuts.  
The photonic electron reconstruction efficiency ($\varepsilon^{\rm reco}$) takes into account track quality cuts applied on the partner electron and the reconstruction cuts on electron pairs.

The photonic electron reconstruction efficiency is estimated by embedding Monte Carlo $\pi^{0}$/$\eta$ and $\gamma$ particles into a full GEANT simulation of the STAR detector. The $\pi^{0}$/$\eta\rightarrow\gamma\gamma$ decays and direct photons are the dominant $\gamma$ sources.
The input spectra of $\pi^0$ in Au+Au collisions at $\sNN=$ 27 and 54.4 GeV analysis are parameterized from $\pi^{0}/\pi^{\pm}$ spectra measurements in Au+Au collisions at $\sqrt{s_{\rm NN}}=$ 39 for the former and 62.4 GeV for the latter \cite{pi0sp,starPIDsp,phenixPIDsp}.
Measurements of direct photon production from Au+Au and p+p collision systems are scaled and combined \cite{dirphosp,Paquet:2015lta,Angelis:1980yc,Angelis:1989zv,Akesson:1989hp}, assuming proportionality to the $N_{\rm coll}\simeq(\frac{dN_{ch}}{d\eta})^{\alpha}+C$ relation where $N_{\rm coll}$ is the number of binary collisions,  $\frac{dN_{ch}}{d\eta}$ is the charged particle multiplicity, $\alpha$ and $C$ are parameters determined from measurements \cite{dirphosp}.
The $\eta$ spectra are scaled from input $\pi^0$ spectra assuming the shapes of their transverse mass $m_{\rm T}$ spectra are the same \cite{etapi0AuAu,STAR:2021tve}. In the simulation, photonic electrons are reconstructed with the same method as in the real data analysis. Figure \ref{fig:dataMC}(a)-(c) show the data and Monte Carlo comparisons of the partner electron $p_{\rm T}$ distribution, the reconstructed pair-DCA and decay-length distributions of dielectrons for the tagged electron with $0.4<p_{\rm T}<2.5$ GeV/$c$ in Au+Au collisions at $\sqrt{s_{\rm NN}}=$ $54.4\,\rm GeV$. The peaks around 5 and 60 cm in Fig.~\ref{fig:dataMC}(c) are caused by photon conversion electrons induced by the beam pipe and the TPC inner field cage (TPC-IFC), respectively, and are well described by the simulation. 
At $p_{\rm T}<0.5\,{\rm GeV}/c$, the photonic electrons are predominately due to Dalitz decays, while at $p_{\rm T}>1.5$ GeV/$c$, electrons from photon conversion in the TPC-IFC become dominant. 
Reconstruction efficiencies for electrons from various sources are combined using their relative contributions to the total photonic electron yields including their $p_{\rm T}$ dependence.
The estimated reconstruction efficiency for $e^{\rm PE}$ in Au+Au collisions at $\sNN=54.4\,{\rm GeV}/c$ is shown as solid circles in Fig. \ref{fig:Eff}(a). Reconstruction efficiencies from various sources are also indicated as dashed lines in this plot. Systematic uncertainties of the $e^{\rm PE}$ reconstruction efficiency are discussed in Sec. \ref{sec:sys}. 
The $e^{\rm PE}$ reconstruction efficiency in 27 GeV is slightly lower than that in 54.4 GeV due to a steeper partner electron $p_{\rm T}$ distribution.

The non-photonic electron to photonic electron yield ratio ($N^{\rm NPE}/N^{\rm PE}$) in Au+Au collisions at $\sNN=$ 27, 54.4, and $200\,\rm GeV$ \cite{starHFe} collisions is shown in Fig. \ref{fig:Eff}(b). Because the charmed hadron production cross section drops faster with the decreasing collision energy than the light hadron production cross section, $N^{\rm NPE}/N^{\rm PE}$ is smaller at lower energies.  The systematic uncertainties of $N^{\rm NPE}/N^{\rm PE}$ in Au+Au collisions include uncertainties propagated from the purities of inclusive electron candidates and photonic electron reconstruction efficiency.

\begin{figure}[htbp]
\centering\includegraphics[width=0.9\linewidth]{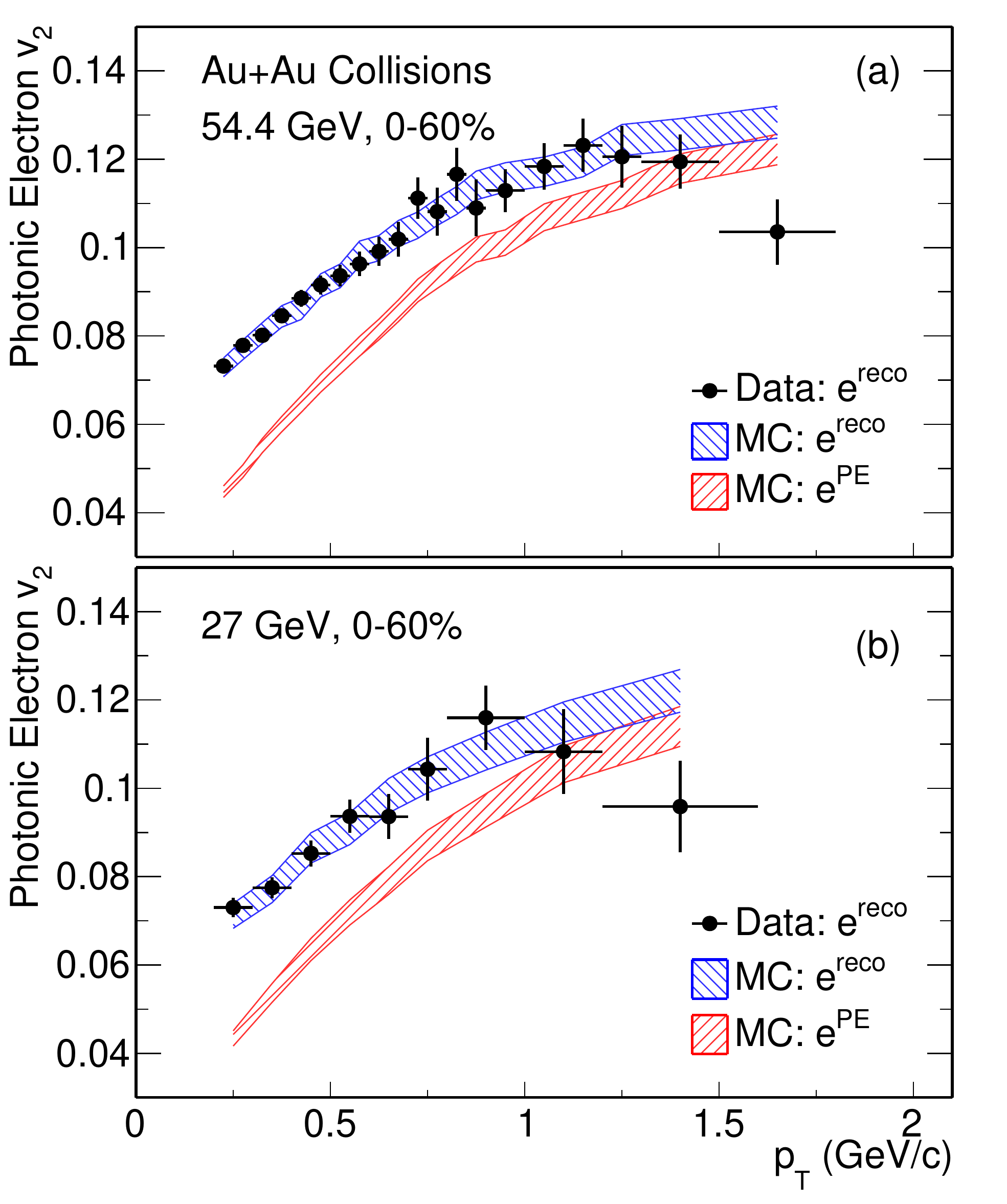}
\caption{Photonic electron $v_2$ distributions from Monte Carlo and real data in Au+Au collisions at $\sqrt{s_{_{\rm NN}}}$ = 54.4 (a) and 27 (b) GeV, respectively. Blue and red bands depict the $v_2$ of reconstructed and total photonic electrons from Monte Carlo, respectively. The black data points are reconstructed photonic electron $v_2$ from real data. The vertical bars denote the statistical uncertainties. The vertical width of blue and red bands are the systematic uncertainties of Monte Carlo $v_2^{\rm reco}$ and $v_2^{\rm PE}$, respectively.}
\label{fig:PEv2}
\end{figure}

The elliptic flow of inclusive electrons ($v_2^{\rm inc}$) is extracted by the event plane $\eta-$sub method \cite{Poskanzer:1998yz}. 
The event plane is reconstructed using TPC tracks at $0.2<p_{\rm T}<2$ GeV/$c$ in the detector's $\eta$ region opposite to that of the electron candidate. An additional $\eta$ gap of $\pm0.05$ is applied between the sub-events to suppress correlations not related to event plane (non-flow effects). Subsequently, $v_2^{\rm inc}$ is calculated as $v_2^{inc} = \langle \cos 2(\phi-\Phi_{\rm EP})\rangle/R$, where $(\phi-\Phi_{\rm EP})$ is the difference in azimuthal angle between electron and the event plane $\Phi_{\rm EP}$ and $R$ is the event plane resolution~\cite{Poskanzer:1998yz,STAR:2016ydv}.
The corrections for the event plane resolution are applied in fine centrality intervals and the average value is found to be $R=0.38$ and $0.44$  in the 0-60\% centrality range in Au+Au $\sNN=$ 27 and 54.4 GeV, respectively.

The $v_2$ of NPE is calculated by:
\begin{equation}
N^{\rm NPE}v_2^{\rm NPE}=N^{\rm inc}v^{\rm inc}_2-N^{\rm PE}v_2^{\rm PE}-\sum_hf_h\cdot N^{\rm inc}v_2^h,
\label{eq:NPEv2}
\end{equation}
where $h$ sums over hadrons ($\pi/p/K$) and $f_h$ are the fractions of hadron contamination in inclusive electrons and their corresponding $v_2^{h}$ are taken from measurements in Au+Au collisions at $\sNN=$ 39 and 62.4 GeV~\cite{Adamczyk:2013gw}. $f_h$ are calculated during the process of electron purity estimation. The $v_2^{\rm PE}$ is $v_2$ of $e^{\rm PE}$ that is estimated with a full detector simulation, similar to that of the $\varepsilon^{\rm reco}$ estimation. The $p_{\rm T}$ and $\phi$ distributions of daughter electrons are weighted according to their parent $p_{\rm T}$ spectra and $v_2$. Due to the absence of published data of $\pi^{0}$ and direct photon from Au+Au collisions at $\sNN=$ 27 and 54.4 GeV, the input $v_2$ of $\pi^0$ and direct photons are scaled from Au+Au at $\sNN=$ 39 and 62.4 GeV \cite{pi0sp, starPIDsp, phenixPIDsp, dirphosp, Angelis:1980yc, Akesson:1989hp, Angelis:1989zv, Paquet:2015lta, dirpho200} measurements. The input $v_2$ of $\eta$ is derived from kaon $v_2$ \cite{Adamczyk:2013gw} at the corresponding energies. 
The simulated $v_2$ for total photonic electron $v_2^{\rm PE}$ are shown with red bands in Fig.~\ref{fig:PEv2}.
The mean $p_{\rm T}$ of parents from reconstructed photonic electrons ($e^{\rm reco}$) is higher compared to parents of total photonic electrons, due to the minimum $p_{\rm T}$ cut on partner electrons. A further consequence of both this and the $p_{\rm T}$ dependence of elliptic flow, is that the $v_{2}$ of $e^{\rm reco}$ ($v_{2}^{\rm reco}$) is larger than $v_{2}^{PE}$ at $p_{\rm T}<2\,{\rm GeV}/c$.
The $v_2^{\rm reco}$ calculated from data and simulation are shown in Fig.~\ref{fig:PEv2}. One can see that $v_2^{\rm reco}$ from simulations in both energies can describe the data very well which validates these simulations. The systematic uncertainties of the photonic electron $v_2$ simulation are evaluated by comparing the difference of $v_{2}^{\rm reco}$ between data and simulation.

In addition to $e^{\rm PE}$, other major background sources are electrons from kaon weak decay ($K_{e3}$) and vector meson decays. 
The relative contributions of $K_{e3}$ and electrons from decayed vector mesons in NPE are estimated using fast simulations assuming that the TPC tracking efficiency is the same for $e^{\rm HF}$ and $K_{e3}$ tracks that satisfy ${\rm DCA}<1.5$ cm. Kaons are decayed by PYTHIA6 \cite{Sjostrand:2006za}, and charged tracks are curved under a magnetic field of B = 0.5 T. The input kaon $p_{\rm T}$ spectrum is taken from $K_S^{0}$ measurements in Au+Au collisions at $\sNN=$ 62.4 \cite{Aggarwal:2010ig} and 27 GeV \cite{STAR:2019bjj}, and kaon $v_2$ is from Au+Au at $\sNN=$ $54.4\,\rm GeV$ measurements. 
Vector meson decay electrons (VM$\rightarrow$e) include $\omega/\rho/\phi\rightarrow e^{+}e^{-}$, $\omega\rightarrow\pi^{0}e^{+}e^{-}$ and $\phi\rightarrow\eta e^{+}e^{-}$. The shape of the vector meson spectra are modified from $\pi^{\pm}$ spectra measured at $\sNN=$ 62.4 and 39 GeV \cite{pi0sp,starPIDsp,phenixPIDsp} assuming that they follow $m_{\rm T}$-scaling \cite{STAR:2021tve}. The $\sNN=$ 39 GeV spectra are scaled to that in $\sNN=$ 27 GeV collisions based on the energy dependence of pion yields measured by STAR \cite{STAR:2017sal}. Their spectra are further normalized based on the measured vector meson to pion yield ratios in $\sNN=$ 200 GeV Au+Au collisions.
The reference $e^{\rm HF}$ yields are first calculated by FONLL (upper limit) \cite{FONLL,Cacciari:1998it} at $\sNN=62.4$ GeV and PYTHIA6 at $\sNN=$ 27 GeV in p+p collisions and then multiplied by the number of binary nucleon-nucleon collisions $N_{\rm coll}$ \cite{Miller:2007ri} and nuclear modification factor $R_{\rm AA}$ \cite{TAMU62v2}. $R_{\rm AA}$ is from model calculations \cite{TAMU62v2} where the evolution of QGP is simulated by the hydrodynamic model. The estimated fractions of the sum of $K_{e3}$ and ${\rm VM}\rightarrow e$ in $e^{\rm NPE}$ is $\sim$$30\%$ and $\sim$$60\%$ at $p_{\rm T}\sim$$0.5$\,GeV/$c$, and decreases to $\sim$$20\%$ and $\sim$$30\%$ at $p_{\rm T}=1.5$\,GeV/$c$ in the $\sNN=$ 54.4 and 27 GeV measurements, respectively. Heavy-flavor decay electron $v_2$ is calculated as:
\begin{equation}
v_2^{\rm HF}=v_2^{\rm NPE}(1+f_{K_{e3}}+f_{VM})-v_2^{K_{e3}}\cdot f_{K_{e3}}-v_2^{VM}\cdot f_{VM},
\label{eq:ke3VMcor}
\end{equation}
where $f_{Ke3}$ and $f_{VM}$ are the estimated yield ratios of $K_{e3}$ and ${\rm VM}\rightarrow e$ to $e^{\rm HF}$ yields in the inclusive electrons, respectively. 
Because the calculated $v_2^{VM}$ and $v_2^{K_{e3}}$ are comparable to $v_2^{\rm NPE}$ in $\sNN=$ 54.4 GeV analysis, the obtained $v_2^{\rm HF}$ differs from $v_2^{\rm NPE}$ by less than 10\%.

The residual non-flow contribution is estimated in the same way as in Ref.~\cite{starHFe} by using $e^{\rm HF}$-hadron correlations in p+p collisions scaled by the hadron multiplicity in Au+Au collisions. The events of p+p collisions are generated by PYTHIA8 \cite{Sjostrand:2007gs} using STAR heavy flavor tune \cite{STAR:2021zvb}. The non-flow contribution to $v_2$ is estimated as:
\begin{equation}
    v_2^{\rm non-flow} = \frac{\left<\sum_{i}\cos2(\phi_e-\phi_i)\right>}{M\left<v_2\right>}.
\end{equation}
 The numerator is from p+p collisions, where $\phi_e$ and $\phi_i$ are the azimuthal angles for $e^{\rm HF}$ and charged hadrons, respectively. 
The summation is over charged hadrons in the same event, and the average is taken over all events. 
 The denominator is from Au+Au collisions, where $M$ is the multiplicity of charged hadrons used for event plane reconstruction and $\left<v_2\right>$ is the corresponding average coefficient of elliptic flow. 
 This estimate is an upper limit of the non-flow effect since possible modifications to jet-like correlations in the hot medium may lead to a reduction in these correlations.

\section{Systematic uncertainties}\label{sec:sys} 
The dominant sources of systematic uncertainties in this analysis include the purity of inclusive electron candidates, the photonic electron reconstruction efficiency, and the photonic electron $v_2$. The systematic uncertainties of inclusive electron candidates purity have been discussed in Section \ref{sec:exp}. The following sources are considered systematic uncertainties of the photonic electron reconstruction efficiency ($\varepsilon^{reco}$): (1) single electron track quality cuts; (2) electron pair reconstruction cuts; (3) the input spectra shapes for $\pi^0/\eta/\gamma$; (4) the estimation of detector material budgets in the simulation. The estimated relative systematic uncertainties of $\varepsilon^{reco}$ are between 3-4\% and 2-6\% in $0.3<p_{\rm T}<2$ GeV/$c$ for $\sNN=$ 27 and 54.4 GeV, respectively. Since both total and reconstructed photonic electron $v_2$ are estimated from the same simulations, the systematic uncertainties of photonic electron $v_2$ are estimated by evaluating the difference of the reconstructed photonic electron $v_2$ between simulation and data shown in Fig. \ref{fig:PEv2}. 
The relative systematic uncertainties of photonic electron $v_2$, estimated by the standard deviation of the relative difference between simulation and data in 0.2$<p_{\rm T}<$1.5~GeV/$c$, are 4\% and 3\% for $\sNN=$ 27 and 54.4 GeV collisions, respectively.
The systematic uncertainties of the fraction of $K_{e3}$ and electrons from vector meson decays in non-photonic electrons are estimated by varying input $e^{\rm HF}$ $R_{\rm AA}$ from using model calculated values \cite{TAMU62v2} to $R_{\rm AA}=1$. The summary of absolute systematic uncertainties from different sources propagated to the $e^{\rm HF}$ $v_2$ are listed in Table \ref{tab:sys}. 
\begin{table*}
\centering
\begin{tabular}{c|c|c|c}
\hline
   \multicolumn{4}{c}{Systematic Uncertainties} \\
\hline
    \multirow{2}*{Sources} & \multicolumn{2}{c|}{Au+Au $\sNN=54.4$\,GeV} & Au+Au $\sNN=27$\,GeV \\
    \cline{2-4}
    ~&$0.35<p_{\rm T}<0.7$ GeV/c & $1.2<p_{\rm T}<1.8$ GeV/c & $0.6<p_{\rm T}<1.6$ GeV/c\\
    \hline
    Electron purity
 & $0.001-0.007$ & $0.001-0.004$ &  $0.006-0.013$ \\
 \hline
    $\varepsilon^{reco}$
 & $0.003-0.023$ & $0.001-0.007$ & $0.021-0.038$  \\
 \hline
    Photonic electron $v_2$
 & $0.017-0.032$ & $0.016-0.018$ & $0.041-0.075$  \\
 \hline
    $K_{e3}$ and vector meson decays
 & negligible & $0.002-0.009$ & $0.001-0.042$\\
 \hline
 Total systematic uncertainties& $0.019-0.040$ & $0.017-0.021$ & $0.071-0.079$\\
 \hline
\end{tabular}
\caption{Summary of absolute systematic uncertainties propagated from various sources to the heavy-flavor decay electron $v_2$.}
\label{tab:sys}
\end{table*}

\begin{figure}[tb]
\centering
\includegraphics[width=0.9\linewidth]{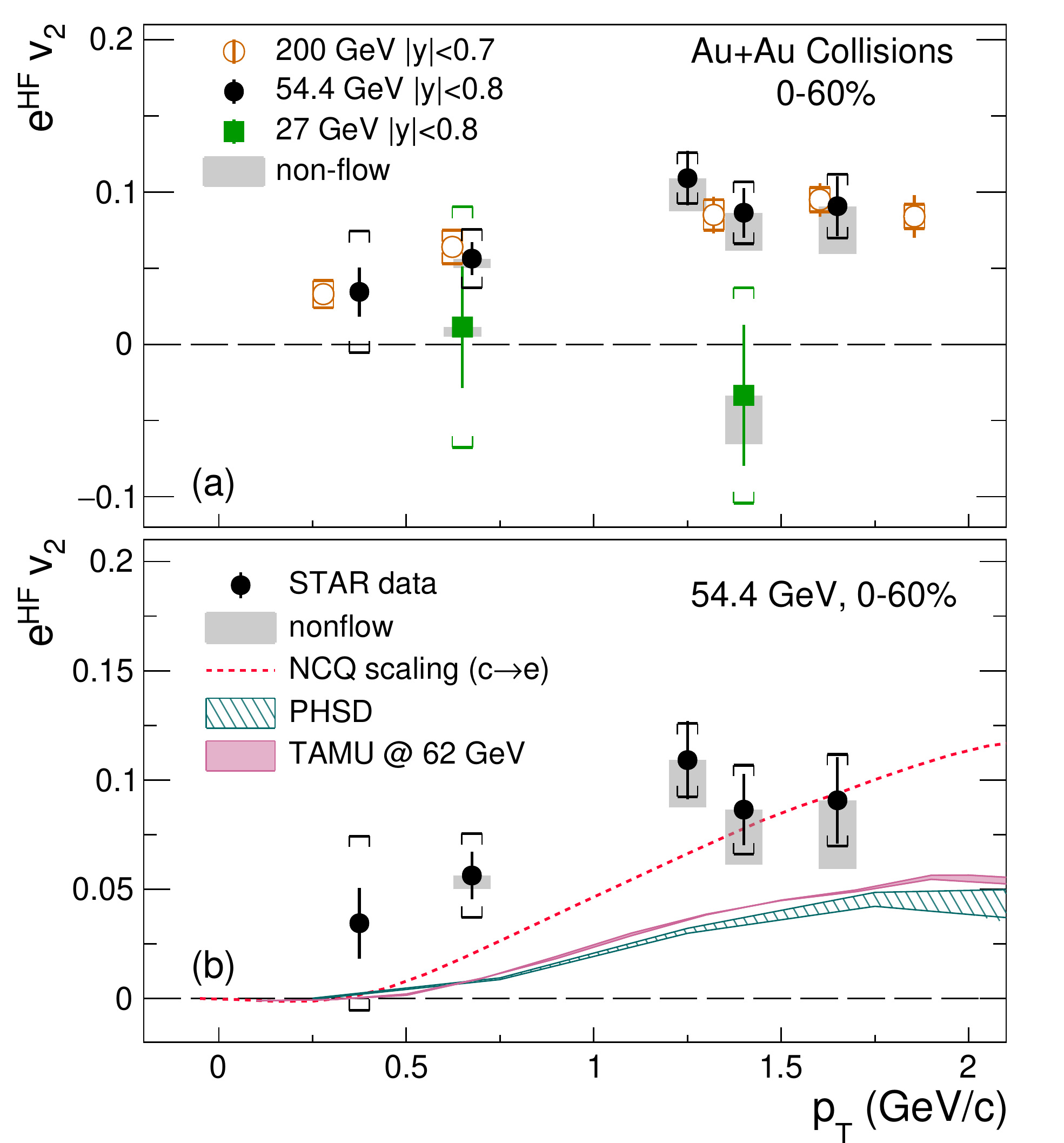}
\caption{(a): Heavy-flavor decay electron $v_2$ as a function of electron $p_{\rm T}$ in Au+Au collisions at $\sqrt{s_{_{\rm NN}}}$ = 54.4 GeV (full circle) and 27 GeV (full square) compared to the previous measurement at $\sNN=$ 200 GeV \cite{starHFe} (open circle). Statistical and systematic uncertainties are shown as error bars and brackets, respectively. Gray boxes indicate the estimated upper limit of non-flow contributions. (b): Heavy-flavor decay electron $v_2$ in Au+Au collisions at $\sNN=$ 54.4 GeV from STAR experiment compared to the TAMU \cite{TAMU62v2} and PHSD \cite{Song:2015sfa,Song:2016rzw} calculations. The dashed line refers to the projected charm-decay electron $v_2$ assuming open charmed hadron $v_2$ follows NCQ scaling with other light hadrons in Au+Au collisions at $\sNN=54.4$ GeV. The $D\rightarrow e$ decay kinematics are simulated in PYTHIA6. The vertical bars and square brackets denote the statistical and systematic uncertainties,  respectively.}
\label{fig:data}
\end{figure}

\section{Results and Discussions}\label{sec:result}
Figure \ref{fig:data}(a) shows elliptic flow $v_2$ of $e^{\rm HF}$ as a function of $p_{\rm T}$ at mid-rapidity ($|y| < 0.8$) in Au+Au collisions at $\sNN=$ 27 and 54.4 GeV from this analysis and those at $\sNN=$ $200\,\rm GeV$ published previously~\cite{starHFe}. The gray hatched area indicates the estimated non-flow contribution to the measured $v_2$ via the event-plane method. Compared to the previous measurements at similar collision energies of $\sNN=$ 39 and $62.4\,\rm GeV$~\cite{starHFe,phenix62HFe}, the results from this analysis are more precise, both in terms of statistical and systematic uncertainties. The $e^{\rm HF}$ $v_2$ in Au+Au $\sNN=$ 54.4 GeV collisions is sizable and is comparable to that at $\sNN=$ 200 GeV collisions in the measured $p_{\rm T}$ region. The integrated $e^{\rm HF}$ $v_2$ within $1.2<p_{\rm T}<2\, {\rm GeV}/c$ is $0.094\,\pm\,0.008\,({\rm stat.})\,\pm \,0.014\,({\rm syst.})$, while the estimated upper limit of non-flow contribution is 0.02.
The significant $v_2$ of $e^{\rm HF}$ observed at $\sNN=$ 54.4\,GeV indicates that charm quarks interact strongly with the QGP medium and may reach local thermal equilibrium in Au+Au collisions at $\sNN=$ 54.4 GeV, even though the collision energy is nearly a factor of 4 lower than $\sNN=$ 200 GeV. The initial energy density at Au+Au $\sNN=$ 200 GeV collisions is about 2 times higher than that of $\sNN=$ $54.4\,\rm GeV$ collisions from a semi-analytical calculation at formation time $\tau_F=0.3$ fm/$c$ \cite{Mendenhall:2020fil}. Consequently, the initial temperature of the QGP medium created in $\sNN=$ $54.4\,\rm GeV$ collisions is lower than that in $\sNN=$ $200\,\rm GeV$ collisions \cite{Rapp:2014hha}. The similar magnitude of $e^{\rm HF}$ $v_2$ between $\sNN=$ 54.4 and $200\,\rm GeV$ collisions suggests that charm quarks gain most collectivity through diffusion inside the QGP medium at the temperature region close to the critical temperature \cite{Adamczyk:2017xur,TAMU62v2}.
The $e^{\rm HF}$ $v_2$ in $\sNN=$ 27 GeV Au+Au collisions are consistent with zero. A smaller charm quark $v_2$ than light quark $v_2$ may hint that charm quarks deviate from local thermal equilibrium; however, the experimental uncertainties are still appreciable.

Figure \ref{fig:data}(b) compares the experimental results of $e^{\rm HF}$ $v_2$ in Au+Au $\sNN=$ 54.4 GeV collisions with two phenomenological model calculations: TAMU \cite{TAMU62v2} and PHSD (parton-hadron string dynamics) \cite{Song:2015sfa,Song:2016rzw}. TAMU calculations are for Au+Au collisions at $\sNN=$ 62 GeV. TAMU and PHSD models assume that the heavy quarks interact with the strongly coupled QCD medium elastically without the gluon radiation process. It is generally accepted that elastic collision scattering should dominate in this low $p_{\rm T}$ region covered by this analysis~\cite{Rapp:2009my}.

In the TAMU model, the microscopic elastic heavy quark interactions with quarks and gluons in the medium are evaluated using non-perturbative $T$-Matrix calculations \cite{He:2011yi, Riek:2010fk}. The calculated heavy quark transport coefficient fed into macroscopic Langevin simulations of heavy quark diffusion through the background medium \cite{He:2011qa, TAMU62v2}.  
The evolution of the QGP is modeled by ideal 2+1D hydrodynamics. 
Heavy quarks hadronize through both coalescence and fragmentation processes. 
In the PHSD model~\cite{Song:2015sfa}, charm quarks interact with the off-shell massive partons in the QGP. The masses and widths of the partons and the scattering cross section are given by the dynamical quasi-particle model which is matched to the lattice QCD equation of state. 
The PHSD model also implements both coalescence and fragmentation mechanism for heavy quark hadronization. 
The hadronized $B$ and $D$ mesons subsequently interact with other hadrons in the hadronic phase with the cross sections calculated from an effective Lagrangian \cite{Song:2015sfa, Song:2016rzw}. 

Both the TAMU and PHSD calculations underestimated measured central $v_2$ values. With the inclusion of the non-flow contribution and uncertainties, model calculations are 1-2$\sigma$ lower than data points at $p_{\rm T}>0.5$ GeV/$c$. A similar observation was found in $D^0$ $v_2$ results at $p_{\rm T}>$ 2.5 GeV/$c$ in $\sNN=$ 200 GeV Au+Au collisions~\cite{Adamczyk:2017xur}. Additionally, neither model takes into account charm baryon contributions which will slightly increase $e^{\rm HF}$ $v_2$ at $p_{\rm T}>1$ GeV/c.

The $e^{\rm HF}$ momentum differs from its parent charm/bottom hadron momentum due to the decay kinematics. In order to compare $v_2$ of charmed hadrons with identified particle $v_2$, a simulation framework is set up to correct for the $p_{\rm T}$ shift from the measured daughter electron to the parent charmed hadrons. The $\Lambda_{c}^{+}$ and $D^0$ are decayed by PYTHIA6 through the semileptonic channel \cite{PDG:2022pth}. 
The nuclear modification factors of charmed hadrons \cite{TAMU62v2} are also included which result in $\sim70\%$ increase in subsquent daughter electrons $v_2$ at $p_{\rm T}\sim0.65$ GeV/$c$. The input charmed hadrons $v_2$ are assumed to follow the number-of-constituent-quark (NCQ) scaling as those of light hadrons in Au+Au collisions at $\sNN=$ 54.4\,GeV \cite{Adamczyk:2015fum, Nayak:2020djj}. Both $\Lambda_{c}^{+}\rightarrow e$ and $D^0\rightarrow e$ are combined according to their decay branching ratios and charmed hadron chemistry measured in $\sNN=$ 200 GeV Au+Au collisions \cite{STAR:2019ank,STAR:2021tte}. The simulated $v_2$ of electrons from charmed hadron decays, shown as the dashed line in Fig.~\ref{fig:data}(b), is consistent with the $e^{\rm HF}$ $v_2$ measured herein. 
This suggests that charmed hadrons obtain significant $v_2$ comparable to those of light hadrons and may be close to thermal equilibrium with the medium in Au+Au collisions at $\sNN=$ 54.4 GeV.

\begin{figure}[htbp]
\centering\includegraphics[width=0.8\linewidth]{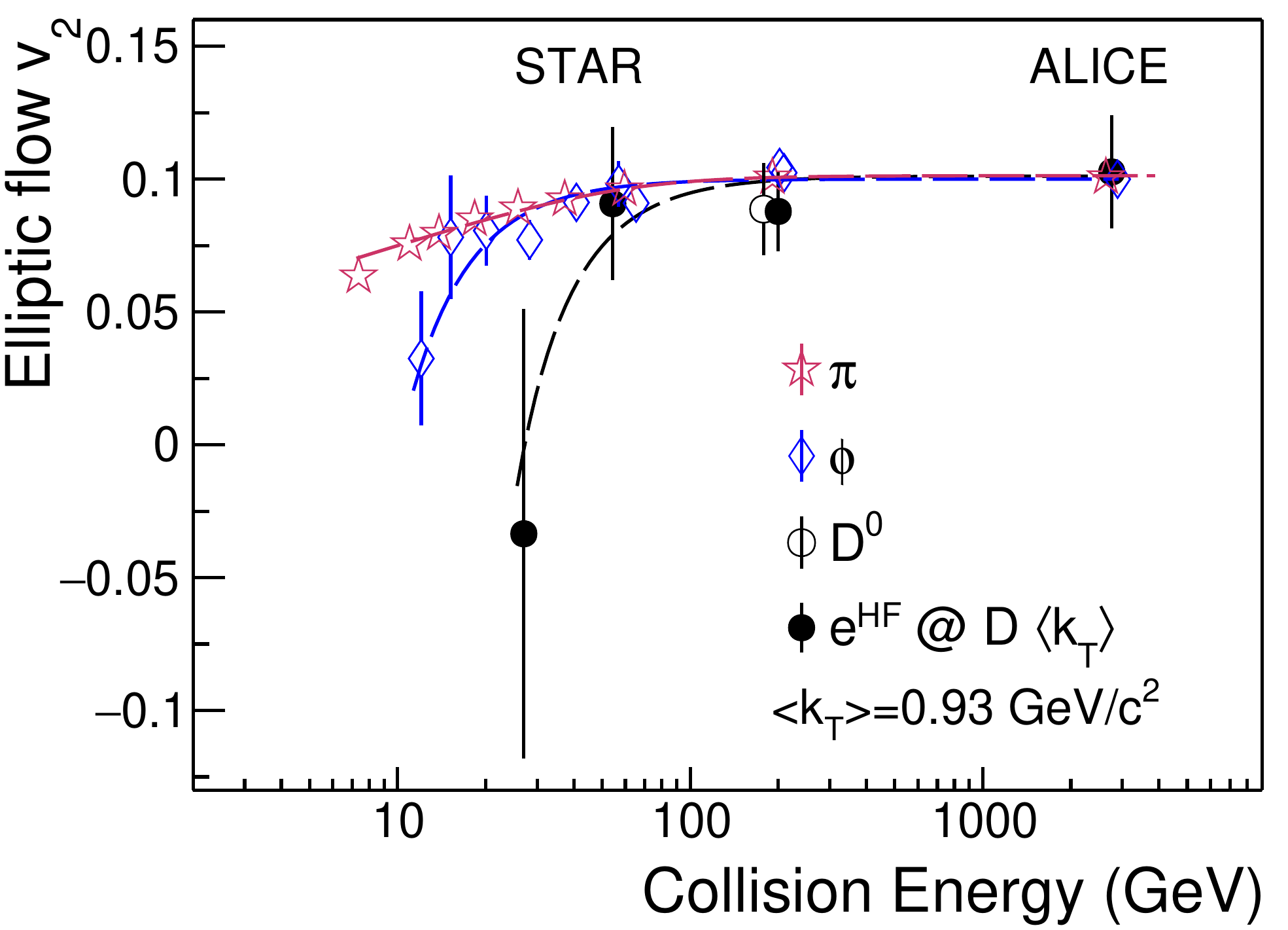}
\caption{Energy dependence of $v_2$ for $\pi^{\pm}$, $\phi$, $D^0$ and $e^{\rm HF}$ at the same transverse mass value $\left<k_{\rm T}\right>=\left<m_{\rm T}-m_0\right>$ = 0.93 GeV/$c^2$. The data points are from or interpolated from STAR \cite{Adamczyk:2013gw,Adams:2003am,STAR:2021twy} and ALICE \cite{Abelev:2014pua,Abelev:2013vea} measurements. The $e^{\rm HF}$ $v_2$ shown here is at the same parent $D^0$ meson transverse mass position using the decay kinematics calculated from PYTHIA6. Data points at the same energy are shifted horizontally for clarity. Error bars depict combined statistical and systematic uncertainties.  The lines are for eye guidance.}
\label{fig:energy}
\end{figure}

Figure \ref{fig:energy} shows the collision energy dependence of $v_2$ for $\pi^{+}$ ($u\bar{d}$), $\phi$($s\bar{s}$), $D^0$ ($c\bar{u}$), and $e^{\rm HF}$ at $\left<k_{\rm T}\right>=\left<m_{\rm T}-m_0\right>=0.93$ GeV/$c^2$. $\phi$ and $D^0$ mesons have smaller scattering cross sections in the hadronic stage, therefore their $v_2$ are sensitive to the early stage dynamics during the fireball evolution. The $e^{\rm HF}$ $v_2$ value is taken at the parent $D^0$ $k_{\rm T}$ value using the decay kinematics calculated by PYTHIA6.
The data points for $\pi^{+}$, $\phi$, and $D^0$ are linearly interpolated from measurements in Au+Au collisions at $\sNN=$ 7.7 - 200\,GeV (0--80\% centrality) \cite{Adamczyk:2013gw,Adams:2003am}, U+U collisions at $\sNN=$ 193 GeV \cite{STAR:2021twy} (0--80\% centrality) and Pb+Pb collisions at $\sNN=$ 2.76\,TeV (0--60\% centrality) \cite{Abelev:2014pua,Abelev:2013vea}. As there are no minimum bias measurements of $e^{\rm HF}$ and $\phi$ $v_2$ in Pb+Pb collisions at $\sNN=2.76$\,TeV, the results from narrower centrality ranges \cite{Abelev:2014pua,Abelev:2013vea} are combined and scaled to $0-60\%$ centrality by eccentricity \cite{Abelev:2013qoq}. The lines in Fig. \ref{fig:energy} are used to guide the eyes.
The $v_2$ of $\phi$, $D^{0}$, and $e^{\rm HF}$ agree with that of $\pi^+$ at top RHIC and LHC energies while deviating from that of $\pi^+$ at low energies. The $v_2$ of $\phi$ is lower than $\pi^+$ $v_2$ at $\sNN=$ 11\,GeV by 1.2$\sigma$, while $e^{\rm HF}$ $v_2$ is $1.3\sigma$ lower than $\phi$ $v_2$ at $\sNN=$ 27\,GeV. A hint of mass hierarchy is observed where the $v_{2}$ of heavier particles drops faster than lighter ones with decreasing collision energy.  This may be suggestive of collision-energy-dependent properties of the QGP. 
Calculations from PHSD \cite{Moreau:2021clr} show that the volume of the QGP and the fraction of energy in the medium to the total collision energy deposited, are smaller at low energy in relation to higher energy collisions; thus, the influence of the QGP medium on final-state particle dynamics is gradually reduced as the collision energies decrease.

\section{Summary}\label{sec:summary}
In summary, new results of heavy-flavor decay electron ($e^{\rm HF}$) elliptic flow $v_2$ at mid-rapidity ($|y|<0.8$) in Au+Au collisions at $\sqrt{s_{_{\rm NN}}}$ = 27 and 54.4\,GeV from STAR are reported. The $e^{\rm HF}$ $v_2$ in Au+Au collisions at $\sNN=27$\,GeV is consistent with zero within large uncertainties, whereas for $\sNN=$ 54.4\,GeV collisions a significant non-zero $v_2$ is observed for $p_{\rm T}<$ 2\,GeV/$c$. The $e^{\rm HF}$ $v_2$ in Au+Au $\sNN=$ 54.4\,GeV is comparable to that at $\sNN=$ $200\,\rm GeV$. TAMU and PHSD transport model calculations underestimate the measured $e^{\rm HF}$ $v_2$ in both $\sNN=$ 200 and 54.4\,GeV at $p_{\rm T}<$ 1\,GeV/$c$. Within the uncertainties, the magnitude of $e^{\rm HF}$ $v_2$ at $\sNN=$ 54.4\,GeV and produced electron $p_{\rm T}>1$ GeV/$c$ is consistent with the scenario that their parent $D$ meson $v_2$ follows the NCQ scaling with light-flavor hadrons in the same collision energy. This suggests that charm quarks gain significant collectivity through the interactions with the expanding QGP medium such that they may reach local thermal equilibrium in Au+Au collisions at $\sNN=$ 54.4 GeV. Our new results are expected to provide new constraints on the charm quark spatial diffusion coefficient, especially its temperature dependence. The energy dependence of measured $v_2$ from various particles ($\pi/\phi/D^0/e^{\rm HF}$) shows a hint of quark-mass dependence. Future measurements on $v_2$ at lower energies, as well as bottom quark $v_2$ results at RHIC and the LHC, will shed new insights into particle collectivity and medium thermalization in heavy-ion collisions.

\section*{Acknowledgements}
We thank the RHIC Operations Group and RCF at BNL, the NERSC Center at LBNL, and the Open Science Grid consortium for providing resources and support.  This work was supported in part by the Office of Nuclear Physics within the U.S. DOE Office of Science, the U.S. National Science Foundation, National Natural Science Foundation of China, Chinese Academy of Science, the Ministry of Science and Technology of China and the Chinese Ministry of Education, the Higher Education Sprout Project by Ministry of Education at NCKU, the National Research Foundation of Korea, Czech Science Foundation and Ministry of Education, Youth and Sports of the Czech Republic, Hungarian National Research, Development and Innovation Office, New National Excellency Programme of the Hungarian Ministry of Human Capacities, Department of Atomic Energy and Department of Science and Technology of the Government of India, the National Science Centre and WUT ID-UB of Poland, the Ministry of Science, Education and Sports of the Republic of Croatia, German Bundesministerium f\"ur Bildung, Wissenschaft, Forschung and Technologie (BMBF), Helmholtz Association, Ministry of Education, Culture, Sports, Science, and Technology (MEXT) and Japan Society for the Promotion of Science (JSPS).




\bibliography{BESNPEv2.bib}

\begin{thebibliography}{80}%
\makeatletter
\providecommand \@ifxundefined [1]{%
 \@ifx{#1\undefined}
}%
\providecommand \@ifnum [1]{%
 \ifnum #1\expandafter \@firstoftwo
 \else \expandafter \@secondoftwo
 \fi
}%
\providecommand \@ifx [1]{%
 \ifx #1\expandafter \@firstoftwo
 \else \expandafter \@secondoftwo
 \fi
}%
\providecommand \natexlab [1]{#1}%
\providecommand \enquote  [1]{``#1''}%
\providecommand \bibnamefont  [1]{#1}%
\providecommand \bibfnamefont [1]{#1}%
\providecommand \citenamefont [1]{#1}%
\providecommand \href@noop [0]{\@secondoftwo}%
\providecommand \href [0]{\begingroup \@sanitize@url \@href}%
\providecommand \@href[1]{\@@startlink{#1}\@@href}%
\providecommand \@@href[1]{\endgroup#1\@@endlink}%
\providecommand \@sanitize@url [0]{\catcode `\\12\catcode `\$12\catcode
  `\&12\catcode `\#12\catcode `\^12\catcode `\_12\catcode `\%12\relax}%
\providecommand \@@startlink[1]{}%
\providecommand \@@endlink[0]{}%
\providecommand \url  [0]{\begingroup\@sanitize@url \@url }%
\providecommand \@url [1]{\endgroup\@href {#1}{\urlprefix }}%
\providecommand \urlprefix  [0]{URL }%
\providecommand \Eprint [0]{\href }%
\providecommand \doibase [0]{http://dx.doi.org/}%
\providecommand \selectlanguage [0]{\@gobble}%
\providecommand \bibinfo  [0]{\@secondoftwo}%
\providecommand \bibfield  [0]{\@secondoftwo}%
\providecommand \translation [1]{[#1]}%
\providecommand \BibitemOpen [0]{}%
\providecommand \bibitemStop [0]{}%
\providecommand \bibitemNoStop [0]{.\EOS\space}%
\providecommand \EOS [0]{\spacefactor3000\relax}%
\providecommand \BibitemShut  [1]{\csname bibitem#1\endcsname}%
\let\auto@bib@innerbib\@empty
\bibitem [{\citenamefont {Adams}\ \emph {et~al.}(2005)\citenamefont {Adams}
  \emph {et~al.}}]{Adams:2005dq}%
  \BibitemOpen
  \bibfield  {author} {\bibinfo {author} {\bibfnamefont {J.}~\bibnamefont
  {Adams}} \emph {et~al.} (\bibinfo {collaboration} {STAR Collaboration}),\
  }\href {\doibase 10.1016/j.nuclphysa.2005.03.085} {\bibfield  {journal}
  {\bibinfo  {journal} {Nucl. Phys. A}\ }\textbf {\bibinfo {volume} {757}},\
  \bibinfo {pages} {102} (\bibinfo {year} {2005})},\ \Eprint
  {http://arxiv.org/abs/nucl-ex/0501009} {arXiv:nucl-ex/0501009} \BibitemShut
  {NoStop}%
\bibitem [{\citenamefont {Adcox}\ \emph {et~al.}(2005)\citenamefont {Adcox}
  \emph {et~al.}}]{Adcox:2004mh}%
  \BibitemOpen
  \bibfield  {author} {\bibinfo {author} {\bibfnamefont {K.}~\bibnamefont
  {Adcox}} \emph {et~al.} (\bibinfo {collaboration} {PHENIX Collaboration}),\
  }\href {\doibase 10.1016/j.nuclphysa.2005.03.086} {\bibfield  {journal}
  {\bibinfo  {journal} {Nucl. Phys. A}\ }\textbf {\bibinfo {volume} {757}},\
  \bibinfo {pages} {184} (\bibinfo {year} {2005})},\ \Eprint
  {http://arxiv.org/abs/nucl-ex/0410003} {arXiv:nucl-ex/0410003} \BibitemShut
  {NoStop}%
\bibitem [{\citenamefont {Muller}\ \emph {et~al.}(2012)\citenamefont {Muller},
  \citenamefont {Schukraft},\ and\ \citenamefont {Wyslouch}}]{Muller:2012zq}%
  \BibitemOpen
  \bibfield  {author} {\bibinfo {author} {\bibfnamefont {B.}~\bibnamefont
  {Muller}}, \bibinfo {author} {\bibfnamefont {J.}~\bibnamefont {Schukraft}}, \
  and\ \bibinfo {author} {\bibfnamefont {B.}~\bibnamefont {Wyslouch}},\ }\href
  {\doibase 10.1146/annurev-nucl-102711-094910} {\bibfield  {journal} {\bibinfo
   {journal} {Ann. Rev. Nucl. Part. Sci.}\ }\textbf {\bibinfo {volume} {62}},\
  \bibinfo {pages} {361} (\bibinfo {year} {2012})},\ \Eprint
  {http://arxiv.org/abs/1202.3233} {arXiv:1202.3233 [hep-ex]} \BibitemShut
  {NoStop}%
\bibitem [{\citenamefont {Svetitsky}(1988)}]{Svetitsky:1987gq}%
  \BibitemOpen
  \bibfield  {author} {\bibinfo {author} {\bibfnamefont {B.}~\bibnamefont
  {Svetitsky}},\ }\href {\doibase 10.1103/PhysRevD.37.2484} {\bibfield
  {journal} {\bibinfo  {journal} {Phys. Rev. D}\ }\textbf {\bibinfo {volume}
  {37}},\ \bibinfo {pages} {2484} (\bibinfo {year} {1988})}\BibitemShut
  {NoStop}%
\bibitem [{\citenamefont {Moore}\ and\ \citenamefont
  {Teaney}(2005)}]{Moore:2004tg}%
  \BibitemOpen
  \bibfield  {author} {\bibinfo {author} {\bibfnamefont {G.~D.}\ \bibnamefont
  {Moore}}\ and\ \bibinfo {author} {\bibfnamefont {D.}~\bibnamefont {Teaney}},\
  }\href {\doibase 10.1103/PhysRevC.71.064904} {\bibfield  {journal} {\bibinfo
  {journal} {Phys. Rev. C}\ }\textbf {\bibinfo {volume} {71}},\ \bibinfo
  {pages} {064904} (\bibinfo {year} {2005})},\ \Eprint
  {http://arxiv.org/abs/hep-ph/0412346} {arXiv:hep-ph/0412346} \BibitemShut
  {NoStop}%
\bibitem [{\citenamefont {Rapp}\ and\ \citenamefont {van
  Hees}(2009)}]{Rapp:2009my}%
  \BibitemOpen
  \bibfield  {author} {\bibinfo {author} {\bibfnamefont {R.}~\bibnamefont
  {Rapp}}\ and\ \bibinfo {author} {\bibfnamefont {H.}~\bibnamefont {van
  Hees}},\ }\href {\doibase 10.1142/9789814293297\_0003} {\  (\bibinfo {year}
  {2009}),\ 10.1142/9789814293297\_0003},\ \Eprint
  {http://arxiv.org/abs/0903.1096} {arXiv:0903.1096 [hep-ph]} \BibitemShut
  {NoStop}%
\bibitem [{\citenamefont {Akiba}\ \emph {et~al.}(2015)\citenamefont {Akiba}
  \emph {et~al.}}]{Akiba:2015jwa}%
  \BibitemOpen
  \bibfield  {author} {\bibinfo {author} {\bibfnamefont {Y.}~\bibnamefont
  {Akiba}} \emph {et~al.},\ }\href@noop {} {\  (\bibinfo {year} {2015})},\
  \Eprint {http://arxiv.org/abs/1502.02730} {arXiv:1502.02730 [nucl-ex]}
  \BibitemShut {NoStop}%
\bibitem [{\citenamefont {Poskanzer}\ and\ \citenamefont
  {Voloshin}(1998)}]{Poskanzer:1998yz}%
  \BibitemOpen
  \bibfield  {author} {\bibinfo {author} {\bibfnamefont {A.~M.}\ \bibnamefont
  {Poskanzer}}\ and\ \bibinfo {author} {\bibfnamefont {S.~A.}\ \bibnamefont
  {Voloshin}},\ }\href {\doibase 10.1103/PhysRevC.58.1671} {\bibfield
  {journal} {\bibinfo  {journal} {Phys. Rev. C}\ }\textbf {\bibinfo {volume}
  {58}},\ \bibinfo {pages} {1671} (\bibinfo {year} {1998})},\ \Eprint
  {http://arxiv.org/abs/nucl-ex/9805001} {arXiv:nucl-ex/9805001} \BibitemShut
  {NoStop}%
\bibitem [{\citenamefont {Voloshin}\ and\ \citenamefont
  {Zhang}(1996)}]{Voloshin:1994mz}%
  \BibitemOpen
  \bibfield  {author} {\bibinfo {author} {\bibfnamefont {S.}~\bibnamefont
  {Voloshin}}\ and\ \bibinfo {author} {\bibfnamefont {Y.}~\bibnamefont
  {Zhang}},\ }\href {\doibase 10.1007/s002880050141} {\bibfield  {journal}
  {\bibinfo  {journal} {Z. Phys. C}\ }\textbf {\bibinfo {volume} {70}},\
  \bibinfo {pages} {665} (\bibinfo {year} {1996})},\ \Eprint
  {http://arxiv.org/abs/hep-ph/9407282} {arXiv:hep-ph/9407282} \BibitemShut
  {NoStop}%
\bibitem [{\citenamefont {Adamczyk}\ \emph
  {et~al.}(2017{\natexlab{a}})\citenamefont {Adamczyk} \emph
  {et~al.}}]{Adamczyk:2017xur}%
  \BibitemOpen
  \bibfield  {author} {\bibinfo {author} {\bibfnamefont {L.}~\bibnamefont
  {Adamczyk}} \emph {et~al.} (\bibinfo {collaboration} {STAR Collaboration}),\
  }\href {\doibase 10.1103/PhysRevLett.118.212301} {\bibfield  {journal}
  {\bibinfo  {journal} {Phys. Rev. Lett.}\ }\textbf {\bibinfo {volume} {118}},\
  \bibinfo {pages} {212301} (\bibinfo {year} {2017}{\natexlab{a}})},\ \Eprint
  {http://arxiv.org/abs/1701.06060} {arXiv:1701.06060 [nucl-ex]} \BibitemShut
  {NoStop}%
\bibitem [{\citenamefont {Acharya}\ \emph
  {et~al.}(2018{\natexlab{a}})\citenamefont {Acharya} \emph
  {et~al.}}]{Acharya:2017qps}%
  \BibitemOpen
  \bibfield  {author} {\bibinfo {author} {\bibfnamefont {S.}~\bibnamefont
  {Acharya}} \emph {et~al.} (\bibinfo {collaboration} {ALICE Collaboration}),\
  }\href {\doibase 10.1103/PhysRevLett.120.102301} {\bibfield  {journal}
  {\bibinfo  {journal} {Phys. Rev. Lett.}\ }\textbf {\bibinfo {volume} {120}},\
  \bibinfo {pages} {102301} (\bibinfo {year} {2018}{\natexlab{a}})},\ \Eprint
  {http://arxiv.org/abs/1707.01005} {arXiv:1707.01005 [nucl-ex]} \BibitemShut
  {NoStop}%
\bibitem [{\citenamefont {Sirunyan}\ \emph
  {et~al.}(2018{\natexlab{a}})\citenamefont {Sirunyan} \emph
  {et~al.}}]{Sirunyan:2017plt}%
  \BibitemOpen
  \bibfield  {author} {\bibinfo {author} {\bibfnamefont {A.~M.}\ \bibnamefont
  {Sirunyan}} \emph {et~al.} (\bibinfo {collaboration} {CMS Collaboration}),\
  }\href {\doibase 10.1103/PhysRevLett.120.202301} {\bibfield  {journal}
  {\bibinfo  {journal} {Phys. Rev. Lett.}\ }\textbf {\bibinfo {volume} {120}},\
  \bibinfo {pages} {202301} (\bibinfo {year} {2018}{\natexlab{a}})},\ \Eprint
  {http://arxiv.org/abs/1708.03497} {arXiv:1708.03497 [nucl-ex]} \BibitemShut
  {NoStop}%
\bibitem [{\citenamefont {Adamczyk}\ \emph {et~al.}(2014)\citenamefont
  {Adamczyk} \emph {et~al.}}]{Adamczyk:2014uip}%
  \BibitemOpen
  \bibfield  {author} {\bibinfo {author} {\bibfnamefont {L.}~\bibnamefont
  {Adamczyk}} \emph {et~al.} (\bibinfo {collaboration} {STAR Collaboration}),\
  }\href {\doibase 10.1103/PhysRevLett.113.142301} {\bibfield  {journal}
  {\bibinfo  {journal} {Phys. Rev. Lett.}\ }\textbf {\bibinfo {volume} {113}},\
  \bibinfo {pages} {142301} (\bibinfo {year} {2014})},\ \bibinfo {note}
  {[Erratum: Phys.Rev.Lett. 121, 229901 (2018)]},\ \Eprint
  {http://arxiv.org/abs/1404.6185} {arXiv:1404.6185 [nucl-ex]} \BibitemShut
  {NoStop}%
\bibitem [{\citenamefont {Adam}\ \emph {et~al.}(2019)\citenamefont {Adam} \emph
  {et~al.}}]{Adam:2018inb}%
  \BibitemOpen
  \bibfield  {author} {\bibinfo {author} {\bibfnamefont {J.}~\bibnamefont
  {Adam}} \emph {et~al.} (\bibinfo {collaboration} {STAR Collaboration}),\
  }\href {\doibase 10.1103/PhysRevC.99.034908} {\bibfield  {journal} {\bibinfo
  {journal} {Phys. Rev. C}\ }\textbf {\bibinfo {volume} {99}},\ \bibinfo
  {pages} {034908} (\bibinfo {year} {2019})},\ \Eprint
  {http://arxiv.org/abs/1812.10224} {arXiv:1812.10224 [nucl-ex]} \BibitemShut
  {NoStop}%
\bibitem [{\citenamefont {Acharya}\ \emph
  {et~al.}(2018{\natexlab{b}})\citenamefont {Acharya} \emph
  {et~al.}}]{Acharya:2018hre}%
  \BibitemOpen
  \bibfield  {author} {\bibinfo {author} {\bibfnamefont {S.}~\bibnamefont
  {Acharya}} \emph {et~al.} (\bibinfo {collaboration} {ALICE Collaboration}),\
  }\href {\doibase 10.1007/JHEP10(2018)174} {\bibfield  {journal} {\bibinfo
  {journal} {JHEP}\ }\textbf {\bibinfo {volume} {10}},\ \bibinfo {pages} {174}
  (\bibinfo {year} {2018}{\natexlab{b}})},\ \Eprint
  {http://arxiv.org/abs/1804.09083} {arXiv:1804.09083 [nucl-ex]} \BibitemShut
  {NoStop}%
\bibitem [{\citenamefont {Sirunyan}\ \emph
  {et~al.}(2018{\natexlab{b}})\citenamefont {Sirunyan} \emph
  {et~al.}}]{Sirunyan:2017xss}%
  \BibitemOpen
  \bibfield  {author} {\bibinfo {author} {\bibfnamefont {A.~M.}\ \bibnamefont
  {Sirunyan}} \emph {et~al.} (\bibinfo {collaboration} {CMS Collaboration}),\
  }\href {\doibase 10.1016/j.physletb.2018.05.074} {\bibfield  {journal}
  {\bibinfo  {journal} {Phys. Lett. B}\ }\textbf {\bibinfo {volume} {782}},\
  \bibinfo {pages} {474} (\bibinfo {year} {2018}{\natexlab{b}})},\ \Eprint
  {http://arxiv.org/abs/1708.04962} {arXiv:1708.04962 [nucl-ex]} \BibitemShut
  {NoStop}%
\bibitem [{\citenamefont {Acharya}\ \emph {et~al.}(2022)\citenamefont {Acharya}
  \emph {et~al.}}]{ALICE:2021rxa}%
  \BibitemOpen
  \bibfield  {author} {\bibinfo {author} {\bibfnamefont {S.}~\bibnamefont
  {Acharya}} \emph {et~al.} (\bibinfo {collaboration} {ALICE}),\ }\href
  {\doibase 10.1007/JHEP01(2022)174} {\bibfield  {journal} {\bibinfo  {journal}
  {JHEP}\ }\textbf {\bibinfo {volume} {01}},\ \bibinfo {pages} {174} (\bibinfo
  {year} {2022})},\ \Eprint {http://arxiv.org/abs/2110.09420} {arXiv:2110.09420
  [nucl-ex]} \BibitemShut {NoStop}%
\bibitem [{\citenamefont {Adamczyk}\ \emph
  {et~al.}(2017{\natexlab{b}})\citenamefont {Adamczyk} \emph
  {et~al.}}]{starHFe}%
  \BibitemOpen
  \bibfield  {author} {\bibinfo {author} {\bibfnamefont {L.}~\bibnamefont
  {Adamczyk}} \emph {et~al.} (\bibinfo {collaboration} {STAR Collaboration}),\
  }\href@noop {} {\bibfield  {journal} {\bibinfo  {journal} {Phys. Rev. C}\
  }\textbf {\bibinfo {volume} {95}},\ \bibinfo {pages} {034907} (\bibinfo
  {year} {2017}{\natexlab{b}})}\BibitemShut {NoStop}%
\bibitem [{\citenamefont {Adare}\ \emph {et~al.}(2007)\citenamefont {Adare}
  \emph {et~al.}}]{Adare:2006nq}%
  \BibitemOpen
  \bibfield  {author} {\bibinfo {author} {\bibfnamefont {A.}~\bibnamefont
  {Adare}} \emph {et~al.} (\bibinfo {collaboration} {PHENIX Collaboration}),\
  }\href {\doibase 10.1103/PhysRevLett.98.172301} {\bibfield  {journal}
  {\bibinfo  {journal} {Phys. Rev. Lett.}\ }\textbf {\bibinfo {volume} {98}},\
  \bibinfo {pages} {172301} (\bibinfo {year} {2007})},\ \Eprint
  {http://arxiv.org/abs/nucl-ex/0611018} {arXiv:nucl-ex/0611018} \BibitemShut
  {NoStop}%
\bibitem [{\citenamefont {Acharya}\ \emph {et~al.}(2020)\citenamefont {Acharya}
  \emph {et~al.}}]{Acharya:2019mom}%
  \BibitemOpen
  \bibfield  {author} {\bibinfo {author} {\bibfnamefont {S.}~\bibnamefont
  {Acharya}} \emph {et~al.} (\bibinfo {collaboration} {ALICE Collaboration}),\
  }\href {\doibase 10.1016/j.physletb.2020.135377} {\bibfield  {journal}
  {\bibinfo  {journal} {Phys. Lett. B}\ }\textbf {\bibinfo {volume} {804}},\
  \bibinfo {pages} {135377} (\bibinfo {year} {2020})},\ \Eprint
  {http://arxiv.org/abs/1910.09110} {arXiv:1910.09110 [nucl-ex]} \BibitemShut
  {NoStop}%
\bibitem [{\citenamefont {Aad}\ \emph {et~al.}(2020)\citenamefont {Aad} \emph
  {et~al.}}]{ATLAS:2020yxw}%
  \BibitemOpen
  \bibfield  {author} {\bibinfo {author} {\bibfnamefont {G.}~\bibnamefont
  {Aad}} \emph {et~al.} (\bibinfo {collaboration} {ATLAS}),\ }\href {\doibase
  10.1016/j.physletb.2020.135595} {\bibfield  {journal} {\bibinfo  {journal}
  {Phys. Lett. B}\ }\textbf {\bibinfo {volume} {807}},\ \bibinfo {pages}
  {135595} (\bibinfo {year} {2020})},\ \Eprint
  {http://arxiv.org/abs/2003.03565} {arXiv:2003.03565 [nucl-ex]} \BibitemShut
  {NoStop}%
\bibitem [{\citenamefont {Beraudo}\ \emph {et~al.}(2018)\citenamefont {Beraudo}
  \emph {et~al.}}]{Rapp:2018qla}%
  \BibitemOpen
  \bibfield  {author} {\bibinfo {author} {\bibfnamefont {A.}~\bibnamefont
  {Beraudo}} \emph {et~al.},\ }\href {\doibase 10.1016/j.nuclphysa.2018.09.002}
  {\bibfield  {journal} {\bibinfo  {journal} {Nucl. Phys. A}\ }\textbf
  {\bibinfo {volume} {979}},\ \bibinfo {pages} {21} (\bibinfo {year} {2018})},\
  \Eprint {http://arxiv.org/abs/1803.03824} {arXiv:1803.03824 [nucl-th]}
  \BibitemShut {NoStop}%
\bibitem [{\citenamefont {Cao}\ \emph {et~al.}(2019)\citenamefont {Cao} \emph
  {et~al.}}]{Cao:2018ews}%
  \BibitemOpen
  \bibfield  {author} {\bibinfo {author} {\bibfnamefont {S.}~\bibnamefont
  {Cao}} \emph {et~al.},\ }\href {\doibase 10.1103/PhysRevC.99.054907}
  {\bibfield  {journal} {\bibinfo  {journal} {Phys. Rev. C}\ }\textbf {\bibinfo
  {volume} {99}},\ \bibinfo {pages} {054907} (\bibinfo {year} {2019})},\
  \Eprint {http://arxiv.org/abs/1809.07894} {arXiv:1809.07894 [nucl-th]}
  \BibitemShut {NoStop}%
\bibitem [{\citenamefont {Dong}\ \emph {et~al.}(2019)\citenamefont {Dong},
  \citenamefont {Lee},\ and\ \citenamefont {Rapp}}]{Dong:2019byy}%
  \BibitemOpen
  \bibfield  {author} {\bibinfo {author} {\bibfnamefont {X.}~\bibnamefont
  {Dong}}, \bibinfo {author} {\bibfnamefont {Y.-J.}\ \bibnamefont {Lee}}, \
  and\ \bibinfo {author} {\bibfnamefont {R.}~\bibnamefont {Rapp}},\ }\href
  {\doibase 10.1146/annurev-nucl-101918-023806} {\bibfield  {journal} {\bibinfo
   {journal} {Ann. Rev. Nucl. Part. Sci.}\ }\textbf {\bibinfo {volume} {69}},\
  \bibinfo {pages} {417} (\bibinfo {year} {2019})},\ \Eprint
  {http://arxiv.org/abs/1903.07709} {arXiv:1903.07709 [nucl-ex]} \BibitemShut
  {NoStop}%
\bibitem [{\citenamefont {Banerjee}\ \emph {et~al.}(2012)\citenamefont
  {Banerjee}, \citenamefont {Datta}, \citenamefont {Gavai},\ and\ \citenamefont
  {Majumdar}}]{Banerjee:2011ra}%
  \BibitemOpen
  \bibfield  {author} {\bibinfo {author} {\bibfnamefont {D.}~\bibnamefont
  {Banerjee}}, \bibinfo {author} {\bibfnamefont {S.}~\bibnamefont {Datta}},
  \bibinfo {author} {\bibfnamefont {R.}~\bibnamefont {Gavai}}, \ and\ \bibinfo
  {author} {\bibfnamefont {P.}~\bibnamefont {Majumdar}},\ }\href {\doibase
  10.1103/PhysRevD.85.014510} {\bibfield  {journal} {\bibinfo  {journal} {Phys.
  Rev. D}\ }\textbf {\bibinfo {volume} {85}},\ \bibinfo {pages} {014510}
  (\bibinfo {year} {2012})},\ \Eprint {http://arxiv.org/abs/1109.5738}
  {arXiv:1109.5738 [hep-lat]} \BibitemShut {NoStop}%
\bibitem [{\citenamefont {Ding}\ \emph {et~al.}(2012)\citenamefont {Ding},
  \citenamefont {Francis}, \citenamefont {Kaczmarek}, \citenamefont {Karsch},
  \citenamefont {Satz},\ and\ \citenamefont {Soeldner}}]{Ding:2012sp}%
  \BibitemOpen
  \bibfield  {author} {\bibinfo {author} {\bibfnamefont {H.~T.}\ \bibnamefont
  {Ding}}, \bibinfo {author} {\bibfnamefont {A.}~\bibnamefont {Francis}},
  \bibinfo {author} {\bibfnamefont {O.}~\bibnamefont {Kaczmarek}}, \bibinfo
  {author} {\bibfnamefont {F.}~\bibnamefont {Karsch}}, \bibinfo {author}
  {\bibfnamefont {H.}~\bibnamefont {Satz}}, \ and\ \bibinfo {author}
  {\bibfnamefont {W.}~\bibnamefont {Soeldner}},\ }\href {\doibase
  10.1103/PhysRevD.86.014509} {\bibfield  {journal} {\bibinfo  {journal} {Phys.
  Rev. D}\ }\textbf {\bibinfo {volume} {86}},\ \bibinfo {pages} {014509}
  (\bibinfo {year} {2012})},\ \Eprint {http://arxiv.org/abs/1204.4945}
  {arXiv:1204.4945 [hep-lat]} \BibitemShut {NoStop}%
\bibitem [{\citenamefont {Brambilla}\ \emph {et~al.}(2020)\citenamefont
  {Brambilla}, \citenamefont {Leino}, \citenamefont {Petreczky},\ and\
  \citenamefont {Vairo}}]{Brambilla:2020siz}%
  \BibitemOpen
  \bibfield  {author} {\bibinfo {author} {\bibfnamefont {N.}~\bibnamefont
  {Brambilla}}, \bibinfo {author} {\bibfnamefont {V.}~\bibnamefont {Leino}},
  \bibinfo {author} {\bibfnamefont {P.}~\bibnamefont {Petreczky}}, \ and\
  \bibinfo {author} {\bibfnamefont {A.}~\bibnamefont {Vairo}},\ }\href
  {\doibase 10.1103/PhysRevD.102.074503} {\bibfield  {journal} {\bibinfo
  {journal} {Phys. Rev. D}\ }\textbf {\bibinfo {volume} {102}},\ \bibinfo
  {pages} {074503} (\bibinfo {year} {2020})},\ \Eprint
  {http://arxiv.org/abs/2007.10078} {arXiv:2007.10078 [hep-lat]} \BibitemShut
  {NoStop}%
\bibitem [{\citenamefont {{A. Adare, C. Aidala, N.N. Ajitanand, and
  others}}(2015)}]{phenix62HFe}%
  \BibitemOpen
  \bibfield  {author} {\bibinfo {author} {\bibnamefont {{A. Adare, C. Aidala,
  N.N. Ajitanand, and others}}} (\bibinfo {collaboration} {PHENIX
  Collaboration}),\ }\href@noop {} {\bibfield  {journal} {\bibinfo  {journal}
  {Phys. Rev. C}\ }\textbf {\bibinfo {volume} {91}},\ \bibinfo {pages} {044907}
  (\bibinfo {year} {2015})}\BibitemShut {NoStop}%
\bibitem [{\citenamefont {Judd}\ \emph {et~al.}(2018)\citenamefont {Judd} \emph
  {et~al.}}]{Judd:2018zbg}%
  \BibitemOpen
  \bibfield  {author} {\bibinfo {author} {\bibfnamefont {E.~G.}\ \bibnamefont
  {Judd}} \emph {et~al.},\ }\href {\doibase 10.1016/j.nima.2018.03.070}
  {\bibfield  {journal} {\bibinfo  {journal} {Nucl. Instrum. Meth. A}\ }\textbf
  {\bibinfo {volume} {902}},\ \bibinfo {pages} {228} (\bibinfo {year}
  {2018})}\BibitemShut {NoStop}%
\bibitem [{\citenamefont {Adler}\ \emph {et~al.}(2001)\citenamefont {Adler},
  \citenamefont {Denisov}, \citenamefont {Garcia}, \citenamefont {Murray},
  \citenamefont {Strobele},\ and\ \citenamefont {White}}]{Adler:2000bd}%
  \BibitemOpen
  \bibfield  {author} {\bibinfo {author} {\bibfnamefont {C.}~\bibnamefont
  {Adler}}, \bibinfo {author} {\bibfnamefont {A.}~\bibnamefont {Denisov}},
  \bibinfo {author} {\bibfnamefont {E.}~\bibnamefont {Garcia}}, \bibinfo
  {author} {\bibfnamefont {M.~J.}\ \bibnamefont {Murray}}, \bibinfo {author}
  {\bibfnamefont {H.}~\bibnamefont {Strobele}}, \ and\ \bibinfo {author}
  {\bibfnamefont {S.~N.}\ \bibnamefont {White}},\ }\href {\doibase
  10.1016/S0168-9002(01)00627-1} {\bibfield  {journal} {\bibinfo  {journal}
  {Nucl. Instrum. Meth. A}\ }\textbf {\bibinfo {volume} {470}},\ \bibinfo
  {pages} {488} (\bibinfo {year} {2001})},\ \Eprint
  {http://arxiv.org/abs/nucl-ex/0008005} {arXiv:nucl-ex/0008005} \BibitemShut
  {NoStop}%
\bibitem [{\citenamefont {Llope}\ \emph {et~al.}(2014)\citenamefont {Llope}
  \emph {et~al.}}]{Llope:2014nva}%
  \BibitemOpen
  \bibfield  {author} {\bibinfo {author} {\bibfnamefont {W.~J.}\ \bibnamefont
  {Llope}} \emph {et~al.},\ }\href {\doibase 10.1016/j.nima.2014.04.080}
  {\bibfield  {journal} {\bibinfo  {journal} {Nucl. Instrum. Meth. A}\ }\textbf
  {\bibinfo {volume} {759}},\ \bibinfo {pages} {23} (\bibinfo {year} {2014})},\
  \Eprint {http://arxiv.org/abs/1403.6855} {arXiv:1403.6855 [physics.ins-det]}
  \BibitemShut {NoStop}%
\bibitem [{\citenamefont {Bonner}\ \emph {et~al.}(2003)\citenamefont {Bonner},
  \citenamefont {Chen}, \citenamefont {Eppley}, \citenamefont {Geurts},
  \citenamefont {Lamas-Valverde}, \citenamefont {Li}, \citenamefont {Llope},
  \citenamefont {Nussbaum}, \citenamefont {Platner},\ and\ \citenamefont
  {Roberts}}]{BONNER2003181}%
  \BibitemOpen
  \bibfield  {author} {\bibinfo {author} {\bibfnamefont {B.}~\bibnamefont
  {Bonner}}, \bibinfo {author} {\bibfnamefont {H.}~\bibnamefont {Chen}},
  \bibinfo {author} {\bibfnamefont {G.}~\bibnamefont {Eppley}}, \bibinfo
  {author} {\bibfnamefont {F.}~\bibnamefont {Geurts}}, \bibinfo {author}
  {\bibfnamefont {J.}~\bibnamefont {Lamas-Valverde}}, \bibinfo {author}
  {\bibfnamefont {C.}~\bibnamefont {Li}}, \bibinfo {author} {\bibfnamefont
  {W.}~\bibnamefont {Llope}}, \bibinfo {author} {\bibfnamefont
  {T.}~\bibnamefont {Nussbaum}}, \bibinfo {author} {\bibfnamefont
  {E.}~\bibnamefont {Platner}}, \ and\ \bibinfo {author} {\bibfnamefont
  {J.}~\bibnamefont {Roberts}},\ }\href {\doibase
  https://doi.org/10.1016/S0168-9002(03)01347-0} {\bibfield  {journal}
  {\bibinfo  {journal} {Nucl. Instrum. Meth. A}\ }\textbf {\bibinfo {volume}
  {508}},\ \bibinfo {pages} {181} (\bibinfo {year} {2003})},\ \bibinfo {note}
  {proceedings of the Sixth International Workshop on Resistive Plate Chambers
  and Related Detectors}\BibitemShut {NoStop}%
\bibitem [{\citenamefont {Miller}\ \emph {et~al.}(2007)\citenamefont {Miller},
  \citenamefont {Reygers}, \citenamefont {Sanders},\ and\ \citenamefont
  {Steinberg}}]{Miller:2007ri}%
  \BibitemOpen
  \bibfield  {author} {\bibinfo {author} {\bibfnamefont {M.~L.}\ \bibnamefont
  {Miller}}, \bibinfo {author} {\bibfnamefont {K.}~\bibnamefont {Reygers}},
  \bibinfo {author} {\bibfnamefont {S.~J.}\ \bibnamefont {Sanders}}, \ and\
  \bibinfo {author} {\bibfnamefont {P.}~\bibnamefont {Steinberg}},\ }\href
  {\doibase 10.1146/annurev.nucl.57.090506.123020} {\bibfield  {journal}
  {\bibinfo  {journal} {Ann. Rev. Nucl. Part. Sci.}\ }\textbf {\bibinfo
  {volume} {57}},\ \bibinfo {pages} {205} (\bibinfo {year} {2007})},\ \Eprint
  {http://arxiv.org/abs/nucl-ex/0701025} {arXiv:nucl-ex/0701025} \BibitemShut
  {NoStop}%
\bibitem [{\citenamefont {Abelev}\ \emph {et~al.}(2010)\citenamefont {Abelev}
  \emph {et~al.}}]{STAR:2009sxc}%
  \BibitemOpen
  \bibfield  {author} {\bibinfo {author} {\bibfnamefont {B.~I.}\ \bibnamefont
  {Abelev}} \emph {et~al.} (\bibinfo {collaboration} {STAR Collaboration}),\
  }\href {\doibase 10.1103/PhysRevC.81.024911} {\bibfield  {journal} {\bibinfo
  {journal} {Phys. Rev. C}\ }\textbf {\bibinfo {volume} {81}},\ \bibinfo
  {pages} {024911} (\bibinfo {year} {2010})},\ \Eprint
  {http://arxiv.org/abs/0909.4131} {arXiv:0909.4131 [nucl-ex]} \BibitemShut
  {NoStop}%
\bibitem [{\citenamefont {Anderson}\ \emph {et~al.}(2003)\citenamefont
  {Anderson} \emph {et~al.}}]{Anderson:2003ur}%
  \BibitemOpen
  \bibfield  {author} {\bibinfo {author} {\bibfnamefont {M.}~\bibnamefont
  {Anderson}} \emph {et~al.},\ }\href {\doibase 10.1016/S0168-9002(02)01964-2}
  {\bibfield  {journal} {\bibinfo  {journal} {Nucl. Instrum. Meth. A}\ }\textbf
  {\bibinfo {volume} {499}},\ \bibinfo {pages} {659} (\bibinfo {year}
  {2003})},\ \Eprint {http://arxiv.org/abs/nucl-ex/0301015}
  {arXiv:nucl-ex/0301015} \BibitemShut {NoStop}%
\bibitem [{\citenamefont {Llope}\ \emph {et~al.}(2004)\citenamefont {Llope}
  \emph {et~al.}}]{Llope:2003ti}%
  \BibitemOpen
  \bibfield  {author} {\bibinfo {author} {\bibfnamefont {W.~J.}\ \bibnamefont
  {Llope}} \emph {et~al.},\ }\href {\doibase 10.1016/j.nima.2003.11.414}
  {\bibfield  {journal} {\bibinfo  {journal} {Nucl. Instrum. Meth. A}\ }\textbf
  {\bibinfo {volume} {522}},\ \bibinfo {pages} {252} (\bibinfo {year}
  {2004})},\ \Eprint {http://arxiv.org/abs/nucl-ex/0308022}
  {arXiv:nucl-ex/0308022} \BibitemShut {NoStop}%
\bibitem [{\citenamefont {Bellwied}\ \emph {et~al.}(2003)\citenamefont
  {Bellwied} \emph {et~al.}}]{STAR:2002bzu}%
  \BibitemOpen
  \bibfield  {author} {\bibinfo {author} {\bibfnamefont {R.}~\bibnamefont
  {Bellwied}} \emph {et~al.} (\bibinfo {collaboration} {STAR Collaboration}),\
  }\href {\doibase 10.1016/S0168-9002(02)01962-9} {\bibfield  {journal}
  {\bibinfo  {journal} {Nucl. Instrum. Meth. A}\ }\textbf {\bibinfo {volume}
  {499}},\ \bibinfo {pages} {640} (\bibinfo {year} {2003})}\BibitemShut
  {NoStop}%
\bibitem [{\citenamefont {Bichsel}(2006)}]{BICHSEL2006154}%
  \BibitemOpen
  \bibfield  {author} {\bibinfo {author} {\bibfnamefont {H.}~\bibnamefont
  {Bichsel}},\ }\href {\doibase https://doi.org/10.1016/j.nima.2006.03.009}
  {\bibfield  {journal} {\bibinfo  {journal} {Nucl. Instrum. Meth. A}\ }\textbf
  {\bibinfo {volume} {562}},\ \bibinfo {pages} {154} (\bibinfo {year}
  {2006})}\BibitemShut {NoStop}%
\bibitem [{\citenamefont {Xu}\ \emph {et~al.}(2010)\citenamefont {Xu} \emph
  {et~al.}}]{XU201028}%
  \BibitemOpen
  \bibfield  {author} {\bibinfo {author} {\bibfnamefont {Y.}~\bibnamefont {Xu}}
  \emph {et~al.},\ }\href {\doibase https://doi.org/10.1016/j.nima.2009.12.011}
  {\bibfield  {journal} {\bibinfo  {journal} {Nucl. Instrum. Meth. A}\ }\textbf
  {\bibinfo {volume} {614}},\ \bibinfo {pages} {28} (\bibinfo {year}
  {2010})}\BibitemShut {NoStop}%
\bibitem [{\citenamefont {Agakishiev}\ \emph {et~al.}(2011)\citenamefont
  {Agakishiev} \emph {et~al.}}]{Agakishiev:2011mr}%
  \BibitemOpen
  \bibfield  {author} {\bibinfo {author} {\bibfnamefont {H.}~\bibnamefont
  {Agakishiev}} \emph {et~al.} (\bibinfo {collaboration} {STAR
  Collaboration}),\ }\href {\doibase 10.1103/PhysRevD.83.052006} {\bibfield
  {journal} {\bibinfo  {journal} {Phys. Rev. D}\ }\textbf {\bibinfo {volume}
  {83}},\ \bibinfo {pages} {052006} (\bibinfo {year} {2011})},\ \Eprint
  {http://arxiv.org/abs/1102.2611} {arXiv:1102.2611 [nucl-ex]} \BibitemShut
  {NoStop}%
\bibitem [{\citenamefont {Adare}\ \emph {et~al.}(2012)\citenamefont {Adare}
  \emph {et~al.}}]{pi0sp}%
  \BibitemOpen
  \bibfield  {author} {\bibinfo {author} {\bibfnamefont {A.}~\bibnamefont
  {Adare}} \emph {et~al.} (\bibinfo {collaboration} {PHENIX Collaboration}),\
  }\href {\doibase 10.1103/PhysRevLett.109.152301} {\bibfield  {journal}
  {\bibinfo  {journal} {Phys. Rev. Lett.}\ }\textbf {\bibinfo {volume} {109}},\
  \bibinfo {pages} {152301} (\bibinfo {year} {2012})}\BibitemShut {NoStop}%
\bibitem [{\citenamefont {Abelev}\ \emph {et~al.}(2009)\citenamefont {Abelev}
  \emph {et~al.}}]{starPIDsp}%
  \BibitemOpen
  \bibfield  {author} {\bibinfo {author} {\bibfnamefont {B.~I.}\ \bibnamefont
  {Abelev}} \emph {et~al.} (\bibinfo {collaboration} {STAR Collaboration}),\
  }\href {\doibase 10.1103/PhysRevC.79.034909} {\bibfield  {journal} {\bibinfo
  {journal} {Phys. Rev. C}\ }\textbf {\bibinfo {volume} {79}},\ \bibinfo
  {pages} {034909} (\bibinfo {year} {2009})},\ \Eprint
  {http://arxiv.org/abs/0808.2041} {arXiv:0808.2041 [nucl-ex]} \BibitemShut
  {NoStop}%
\bibitem [{\citenamefont {Abelev}\ \emph {et~al.}(2007)\citenamefont {Abelev},
  \citenamefont {Aggarwal}, \citenamefont {Ahammed}, \citenamefont {Anderson},\
  and\ \citenamefont {Arkhipkin}}]{phenixPIDsp}%
  \BibitemOpen
  \bibfield  {author} {\bibinfo {author} {\bibfnamefont {B.}~\bibnamefont
  {Abelev}}, \bibinfo {author} {\bibfnamefont {M.}~\bibnamefont {Aggarwal}},
  \bibinfo {author} {\bibfnamefont {Z.}~\bibnamefont {Ahammed}}, \bibinfo
  {author} {\bibfnamefont {B.}~\bibnamefont {Anderson}}, \ and\ \bibinfo
  {author} {\bibfnamefont {D.}~\bibnamefont {Arkhipkin}},\ }\href@noop {}
  {\bibfield  {journal} {\bibinfo  {journal} {Physics Letters B}\ }\textbf
  {\bibinfo {volume} {655}},\ \bibinfo {pages} {104 } (\bibinfo {year}
  {2007})}\BibitemShut {NoStop}%
\bibitem [{\citenamefont {Adare}\ \emph {et~al.}(2019)\citenamefont {Adare}
  \emph {et~al.}}]{dirphosp}%
  \BibitemOpen
  \bibfield  {author} {\bibinfo {author} {\bibfnamefont {A.}~\bibnamefont
  {Adare}} \emph {et~al.} (\bibinfo {collaboration} {PHENIX Collaboration}),\
  }\href {\doibase 10.1103/PhysRevLett.123.022301} {\bibfield  {journal}
  {\bibinfo  {journal} {Phys. Rev. Lett.}\ }\textbf {\bibinfo {volume} {123}},\
  \bibinfo {pages} {022301} (\bibinfo {year} {2019})}\BibitemShut {NoStop}%
\bibitem [{\citenamefont {Paquet}\ \emph {et~al.}(2016)\citenamefont {Paquet},
  \citenamefont {Shen}, \citenamefont {Denicol}, \citenamefont {Luzum},
  \citenamefont {Schenke}, \citenamefont {Jeon},\ and\ \citenamefont
  {Gale}}]{Paquet:2015lta}%
  \BibitemOpen
  \bibfield  {author} {\bibinfo {author} {\bibfnamefont {J.-F.}\ \bibnamefont
  {Paquet}}, \bibinfo {author} {\bibfnamefont {C.}~\bibnamefont {Shen}},
  \bibinfo {author} {\bibfnamefont {G.~S.}\ \bibnamefont {Denicol}}, \bibinfo
  {author} {\bibfnamefont {M.}~\bibnamefont {Luzum}}, \bibinfo {author}
  {\bibfnamefont {B.}~\bibnamefont {Schenke}}, \bibinfo {author} {\bibfnamefont
  {S.}~\bibnamefont {Jeon}}, \ and\ \bibinfo {author} {\bibfnamefont
  {C.}~\bibnamefont {Gale}},\ }\href {\doibase 10.1103/PhysRevC.93.044906}
  {\bibfield  {journal} {\bibinfo  {journal} {Phys. Rev. C}\ }\textbf {\bibinfo
  {volume} {93}},\ \bibinfo {pages} {044906} (\bibinfo {year} {2016})},\
  \Eprint {http://arxiv.org/abs/1509.06738} {arXiv:1509.06738 [hep-ph]}
  \BibitemShut {NoStop}%
\bibitem [{\citenamefont {Angelis}\ \emph {et~al.}(1980)\citenamefont {Angelis}
  \emph {et~al.}}]{Angelis:1980yc}%
  \BibitemOpen
  \bibfield  {author} {\bibinfo {author} {\bibfnamefont {A.}~\bibnamefont
  {Angelis}} \emph {et~al.} (\bibinfo {collaboration}
  {CERN-Columbia-Oxford-Rockefeller, CCOR}),\ }\href {\doibase
  10.1016/0370-2693(80)90836-9} {\bibfield  {journal} {\bibinfo  {journal}
  {Phys. Lett. B}\ }\textbf {\bibinfo {volume} {94}},\ \bibinfo {pages} {106}
  (\bibinfo {year} {1980})}\BibitemShut {NoStop}%
\bibitem [{\citenamefont {Angelis}\ \emph {et~al.}(1989)\citenamefont {Angelis}
  \emph {et~al.}}]{Angelis:1989zv}%
  \BibitemOpen
  \bibfield  {author} {\bibinfo {author} {\bibfnamefont {A.}~\bibnamefont
  {Angelis}} \emph {et~al.} (\bibinfo {collaboration} {CMOR}),\ }\href
  {\doibase 10.1016/0550-3213(89)90305-2} {\bibfield  {journal} {\bibinfo
  {journal} {Nucl. Phys. B}\ }\textbf {\bibinfo {volume} {327}},\ \bibinfo
  {pages} {541} (\bibinfo {year} {1989})}\BibitemShut {NoStop}%
\bibitem [{\citenamefont {Akesson}\ \emph {et~al.}(1990)\citenamefont {Akesson}
  \emph {et~al.}}]{Akesson:1989hp}%
  \BibitemOpen
  \bibfield  {author} {\bibinfo {author} {\bibfnamefont {T.}~\bibnamefont
  {Akesson}} \emph {et~al.} (\bibinfo {collaboration} {Axial Field
  Spectrometer}),\ }\href@noop {} {\bibfield  {journal} {\bibinfo  {journal}
  {Sov. J. Nucl. Phys.}\ }\textbf {\bibinfo {volume} {51}},\ \bibinfo {pages}
  {836} (\bibinfo {year} {1990})}\BibitemShut {NoStop}%
\bibitem [{\citenamefont {Adare}\ \emph {et~al.}(2013)\citenamefont {Adare},
  \citenamefont {Afanasiev},\ and\ \citenamefont {Aidala}}]{etapi0AuAu}%
  \BibitemOpen
  \bibfield  {author} {\bibinfo {author} {\bibfnamefont {A.}~\bibnamefont
  {Adare}}, \bibinfo {author} {\bibfnamefont {S.}~\bibnamefont {Afanasiev}}, \
  and\ \bibinfo {author} {\bibfnamefont {C.}~\bibnamefont {Aidala}} (\bibinfo
  {collaboration} {PHENIX Collaboration}),\ }\href {\doibase
  10.1103/PhysRevC.87.034911} {\bibfield  {journal} {\bibinfo  {journal} {Phys.
  Rev. C}\ }\textbf {\bibinfo {volume} {87}},\ \bibinfo {pages} {034911}
  (\bibinfo {year} {2013})}\BibitemShut {NoStop}%
\bibitem [{\citenamefont {Abdallah}\ \emph
  {et~al.}(2022{\natexlab{a}})\citenamefont {Abdallah} \emph
  {et~al.}}]{STAR:2021tve}%
  \BibitemOpen
  \bibfield  {author} {\bibinfo {author} {\bibfnamefont {M.}~\bibnamefont
  {Abdallah}} \emph {et~al.} (\bibinfo {collaboration} {STAR Collaboration}),\
  }\href {\doibase 10.1103/PhysRevD.105.032007} {\bibfield  {journal} {\bibinfo
   {journal} {Phys. Rev. D}\ }\textbf {\bibinfo {volume} {105}},\ \bibinfo
  {pages} {032007} (\bibinfo {year} {2022}{\natexlab{a}})},\ \Eprint
  {http://arxiv.org/abs/2109.13191} {arXiv:2109.13191 [nucl-ex]} \BibitemShut
  {NoStop}%
\bibitem [{\citenamefont {Adamczyk}\ \emph
  {et~al.}(2016{\natexlab{a}})\citenamefont {Adamczyk} \emph
  {et~al.}}]{STAR:2016ydv}%
  \BibitemOpen
  \bibfield  {author} {\bibinfo {author} {\bibfnamefont {L.}~\bibnamefont
  {Adamczyk}} \emph {et~al.} (\bibinfo {collaboration} {STAR Collaboration}),\
  }\href {\doibase 10.1103/PhysRevC.94.034908} {\bibfield  {journal} {\bibinfo
  {journal} {Phys. Rev. C}\ }\textbf {\bibinfo {volume} {94}},\ \bibinfo
  {pages} {034908} (\bibinfo {year} {2016}{\natexlab{a}})},\ \Eprint
  {http://arxiv.org/abs/1601.07052} {arXiv:1601.07052 [nucl-ex]} \BibitemShut
  {NoStop}%
\bibitem [{\citenamefont {Adamczyk}\ \emph {et~al.}(2013)\citenamefont
  {Adamczyk} \emph {et~al.}}]{Adamczyk:2013gw}%
  \BibitemOpen
  \bibfield  {author} {\bibinfo {author} {\bibfnamefont {L.}~\bibnamefont
  {Adamczyk}} \emph {et~al.} (\bibinfo {collaboration} {STAR Collaboration}),\
  }\href {\doibase 10.1103/PhysRevC.88.014902} {\bibfield  {journal} {\bibinfo
  {journal} {Phys. Rev. C}\ }\textbf {\bibinfo {volume} {88}},\ \bibinfo
  {pages} {014902} (\bibinfo {year} {2013})},\ \Eprint
  {http://arxiv.org/abs/1301.2348} {arXiv:1301.2348 [nucl-ex]} \BibitemShut
  {NoStop}%
\bibitem [{\citenamefont {Adare}\ \emph {et~al.}(2015)\citenamefont {Adare}
  \emph {et~al.}}]{dirpho200}%
  \BibitemOpen
  \bibfield  {author} {\bibinfo {author} {\bibfnamefont {A.}~\bibnamefont
  {Adare}} \emph {et~al.} (\bibinfo {collaboration} {PHENIX Collaboration}),\
  }\href@noop {} {\bibfield  {journal} {\bibinfo  {journal} {Phys. Rev. C}\
  }\textbf {\bibinfo {volume} {91}},\ \bibinfo {pages} {064904} (\bibinfo
  {year} {2015})}\BibitemShut {NoStop}%
\bibitem [{\citenamefont {Sjostrand}\ \emph {et~al.}(2006)\citenamefont
  {Sjostrand}, \citenamefont {Mrenna},\ and\ \citenamefont
  {Skands}}]{Sjostrand:2006za}%
  \BibitemOpen
  \bibfield  {author} {\bibinfo {author} {\bibfnamefont {T.}~\bibnamefont
  {Sjostrand}}, \bibinfo {author} {\bibfnamefont {S.}~\bibnamefont {Mrenna}}, \
  and\ \bibinfo {author} {\bibfnamefont {P.~Z.}\ \bibnamefont {Skands}},\
  }\href {\doibase 10.1088/1126-6708/2006/05/026} {\bibfield  {journal}
  {\bibinfo  {journal} {JHEP}\ }\textbf {\bibinfo {volume} {05}},\ \bibinfo
  {pages} {026} (\bibinfo {year} {2006})},\ \Eprint
  {http://arxiv.org/abs/hep-ph/0603175} {arXiv:hep-ph/0603175} \BibitemShut
  {NoStop}%
\bibitem [{\citenamefont {Aggarwal}\ \emph {et~al.}(2011)\citenamefont
  {Aggarwal} \emph {et~al.}}]{Aggarwal:2010ig}%
  \BibitemOpen
  \bibfield  {author} {\bibinfo {author} {\bibfnamefont {M.~M.}\ \bibnamefont
  {Aggarwal}} \emph {et~al.} (\bibinfo {collaboration} {STAR Collaboration}),\
  }\href {\doibase 10.1103/PhysRevC.83.024901} {\bibfield  {journal} {\bibinfo
  {journal} {Phys. Rev. C}\ }\textbf {\bibinfo {volume} {83}},\ \bibinfo
  {pages} {024901} (\bibinfo {year} {2011})},\ \Eprint
  {http://arxiv.org/abs/1010.0142} {arXiv:1010.0142 [nucl-ex]} \BibitemShut
  {NoStop}%
\bibitem [{\citenamefont {Adam}\ \emph
  {et~al.}(2020{\natexlab{a}})\citenamefont {Adam} \emph
  {et~al.}}]{STAR:2019bjj}%
  \BibitemOpen
  \bibfield  {author} {\bibinfo {author} {\bibfnamefont {J.}~\bibnamefont
  {Adam}} \emph {et~al.} (\bibinfo {collaboration} {STAR Collaboration}),\
  }\href {\doibase 10.1103/PhysRevC.102.034909} {\bibfield  {journal} {\bibinfo
   {journal} {Phys. Rev. C}\ }\textbf {\bibinfo {volume} {102}},\ \bibinfo
  {pages} {034909} (\bibinfo {year} {2020}{\natexlab{a}})},\ \Eprint
  {http://arxiv.org/abs/1906.03732} {arXiv:1906.03732 [nucl-ex]} \BibitemShut
  {NoStop}%
\bibitem [{\citenamefont {Adamczyk}\ \emph
  {et~al.}(2017{\natexlab{c}})\citenamefont {Adamczyk} \emph
  {et~al.}}]{STAR:2017sal}%
  \BibitemOpen
  \bibfield  {author} {\bibinfo {author} {\bibfnamefont {L.}~\bibnamefont
  {Adamczyk}} \emph {et~al.} (\bibinfo {collaboration} {STAR Collaboration}),\
  }\href {\doibase 10.1103/PhysRevC.96.044904} {\bibfield  {journal} {\bibinfo
  {journal} {Phys. Rev. C}\ }\textbf {\bibinfo {volume} {96}},\ \bibinfo
  {pages} {044904} (\bibinfo {year} {2017}{\natexlab{c}})},\ \Eprint
  {http://arxiv.org/abs/1701.07065} {arXiv:1701.07065 [nucl-ex]} \BibitemShut
  {NoStop}%
\bibitem [{\citenamefont {Cacciari}\ \emph {et~al.}(2001)\citenamefont
  {Cacciari}, \citenamefont {Frixione},\ and\ \citenamefont {Nason}}]{FONLL}%
  \BibitemOpen
  \bibfield  {author} {\bibinfo {author} {\bibfnamefont {M.}~\bibnamefont
  {Cacciari}}, \bibinfo {author} {\bibfnamefont {S.}~\bibnamefont {Frixione}},
  \ and\ \bibinfo {author} {\bibfnamefont {P.}~\bibnamefont {Nason}},\ }\href
  {\doibase 10.1088/1126-6708/2001/03/006} {\bibfield  {journal} {\bibinfo
  {journal} {JHEP}\ }\textbf {\bibinfo {volume} {03}},\ \bibinfo {pages} {006}
  (\bibinfo {year} {2001})},\ \Eprint {http://arxiv.org/abs/hep-ph/0102134}
  {arXiv:hep-ph/0102134} \BibitemShut {NoStop}%
\bibitem [{\citenamefont {Cacciari}\ \emph {et~al.}(1998)\citenamefont
  {Cacciari}, \citenamefont {Greco},\ and\ \citenamefont
  {Nason}}]{Cacciari:1998it}%
  \BibitemOpen
  \bibfield  {author} {\bibinfo {author} {\bibfnamefont {M.}~\bibnamefont
  {Cacciari}}, \bibinfo {author} {\bibfnamefont {M.}~\bibnamefont {Greco}}, \
  and\ \bibinfo {author} {\bibfnamefont {P.}~\bibnamefont {Nason}},\ }\href
  {\doibase 10.1088/1126-6708/1998/05/007} {\bibfield  {journal} {\bibinfo
  {journal} {JHEP}\ }\textbf {\bibinfo {volume} {05}},\ \bibinfo {pages} {007}
  (\bibinfo {year} {1998})},\ \Eprint {http://arxiv.org/abs/hep-ph/9803400}
  {arXiv:hep-ph/9803400} \BibitemShut {NoStop}%
\bibitem [{\citenamefont {He}\ \emph {et~al.}(2015)\citenamefont {He},
  \citenamefont {Fries},\ and\ \citenamefont {Rapp}}]{TAMU62v2}%
  \BibitemOpen
  \bibfield  {author} {\bibinfo {author} {\bibfnamefont {M.}~\bibnamefont
  {He}}, \bibinfo {author} {\bibfnamefont {R.~J.}\ \bibnamefont {Fries}}, \
  and\ \bibinfo {author} {\bibfnamefont {R.}~\bibnamefont {Rapp}},\ }\href
  {\doibase 10.1103/PhysRevC.91.024904} {\bibfield  {journal} {\bibinfo
  {journal} {Phys. Rev. C}\ }\textbf {\bibinfo {volume} {91}},\ \bibinfo
  {pages} {024904} (\bibinfo {year} {2015})}\BibitemShut {NoStop}%
\bibitem [{\citenamefont {Sjostrand}\ \emph {et~al.}(2008)\citenamefont
  {Sjostrand}, \citenamefont {Mrenna},\ and\ \citenamefont
  {Skands}}]{Sjostrand:2007gs}%
  \BibitemOpen
  \bibfield  {author} {\bibinfo {author} {\bibfnamefont {T.}~\bibnamefont
  {Sjostrand}}, \bibinfo {author} {\bibfnamefont {S.}~\bibnamefont {Mrenna}}, \
  and\ \bibinfo {author} {\bibfnamefont {P.~Z.}\ \bibnamefont {Skands}},\
  }\href {\doibase 10.1016/j.cpc.2008.01.036} {\bibfield  {journal} {\bibinfo
  {journal} {Comput. Phys. Commun.}\ }\textbf {\bibinfo {volume} {178}},\
  \bibinfo {pages} {852} (\bibinfo {year} {2008})},\ \Eprint
  {http://arxiv.org/abs/0710.3820} {arXiv:0710.3820 [hep-ph]} \BibitemShut
  {NoStop}%
\bibitem [{\citenamefont {Abdallah}\ \emph
  {et~al.}(2022{\natexlab{b}})\citenamefont {Abdallah} \emph
  {et~al.}}]{STAR:2021zvb}%
  \BibitemOpen
  \bibfield  {author} {\bibinfo {author} {\bibfnamefont {M.}~\bibnamefont
  {Abdallah}} \emph {et~al.} (\bibinfo {collaboration} {STAR Collaboration}),\
  }\href {\doibase 10.1016/j.physletb.2021.136865} {\bibfield  {journal}
  {\bibinfo  {journal} {Phys. Lett. B}\ }\textbf {\bibinfo {volume} {825}},\
  \bibinfo {pages} {136865} (\bibinfo {year} {2022}{\natexlab{b}})},\ \Eprint
  {http://arxiv.org/abs/2110.09666} {arXiv:2110.09666 [nucl-ex]} \BibitemShut
  {NoStop}%
\bibitem [{\citenamefont {Song}\ \emph {et~al.}(2015)\citenamefont {Song},
  \citenamefont {Berrehrah}, \citenamefont {Cabrera}, \citenamefont
  {Torres-Rincon}, \citenamefont {Tolos}, \citenamefont {Cassing},\ and\
  \citenamefont {Bratkovskaya}}]{Song:2015sfa}%
  \BibitemOpen
  \bibfield  {author} {\bibinfo {author} {\bibfnamefont {T.}~\bibnamefont
  {Song}}, \bibinfo {author} {\bibfnamefont {H.}~\bibnamefont {Berrehrah}},
  \bibinfo {author} {\bibfnamefont {D.}~\bibnamefont {Cabrera}}, \bibinfo
  {author} {\bibfnamefont {J.~M.}\ \bibnamefont {Torres-Rincon}}, \bibinfo
  {author} {\bibfnamefont {L.}~\bibnamefont {Tolos}}, \bibinfo {author}
  {\bibfnamefont {W.}~\bibnamefont {Cassing}}, \ and\ \bibinfo {author}
  {\bibfnamefont {E.}~\bibnamefont {Bratkovskaya}},\ }\href {\doibase
  10.1103/PhysRevC.92.014910} {\bibfield  {journal} {\bibinfo  {journal} {Phys.
  Rev. C}\ }\textbf {\bibinfo {volume} {92}},\ \bibinfo {pages} {014910}
  (\bibinfo {year} {2015})},\ \Eprint {http://arxiv.org/abs/1503.03039}
  {arXiv:1503.03039 [nucl-th]} \BibitemShut {NoStop}%
\bibitem [{\citenamefont {Song}\ \emph {et~al.}(2017)\citenamefont {Song},
  \citenamefont {Berrehrah}, \citenamefont {Torres-Rincon}, \citenamefont
  {Tolos}, \citenamefont {Cabrera}, \citenamefont {Cassing},\ and\
  \citenamefont {Bratkovskaya}}]{Song:2016rzw}%
  \BibitemOpen
  \bibfield  {author} {\bibinfo {author} {\bibfnamefont {T.}~\bibnamefont
  {Song}}, \bibinfo {author} {\bibfnamefont {H.}~\bibnamefont {Berrehrah}},
  \bibinfo {author} {\bibfnamefont {J.~M.}\ \bibnamefont {Torres-Rincon}},
  \bibinfo {author} {\bibfnamefont {L.}~\bibnamefont {Tolos}}, \bibinfo
  {author} {\bibfnamefont {D.}~\bibnamefont {Cabrera}}, \bibinfo {author}
  {\bibfnamefont {W.}~\bibnamefont {Cassing}}, \ and\ \bibinfo {author}
  {\bibfnamefont {E.}~\bibnamefont {Bratkovskaya}},\ }\href {\doibase
  10.1103/PhysRevC.96.014905} {\bibfield  {journal} {\bibinfo  {journal} {Phys.
  Rev. C}\ }\textbf {\bibinfo {volume} {96}},\ \bibinfo {pages} {014905}
  (\bibinfo {year} {2017})},\ \Eprint {http://arxiv.org/abs/1605.07887}
  {arXiv:1605.07887 [nucl-th]} \BibitemShut {NoStop}%
\bibitem [{\citenamefont {Mendenhall}\ and\ \citenamefont
  {Lin}(2021)}]{Mendenhall:2020fil}%
  \BibitemOpen
  \bibfield  {author} {\bibinfo {author} {\bibfnamefont {T.}~\bibnamefont
  {Mendenhall}}\ and\ \bibinfo {author} {\bibfnamefont {Z.-W.}\ \bibnamefont
  {Lin}},\ }\href {\doibase 10.1103/PhysRevC.103.024907} {\bibfield  {journal}
  {\bibinfo  {journal} {Phys. Rev. C}\ }\textbf {\bibinfo {volume} {103}},\
  \bibinfo {pages} {024907} (\bibinfo {year} {2021})},\ \Eprint
  {http://arxiv.org/abs/2012.13825} {arXiv:2012.13825 [nucl-th]} \BibitemShut
  {NoStop}%
\bibitem [{\citenamefont {Rapp}\ and\ \citenamefont {van
  Hees}(2016)}]{Rapp:2014hha}%
  \BibitemOpen
  \bibfield  {author} {\bibinfo {author} {\bibfnamefont {R.}~\bibnamefont
  {Rapp}}\ and\ \bibinfo {author} {\bibfnamefont {H.}~\bibnamefont {van
  Hees}},\ }\href {\doibase 10.1016/j.physletb.2015.12.065} {\bibfield
  {journal} {\bibinfo  {journal} {Phys. Lett. B}\ }\textbf {\bibinfo {volume}
  {753}},\ \bibinfo {pages} {586} (\bibinfo {year} {2016})},\ \Eprint
  {http://arxiv.org/abs/1411.4612} {arXiv:1411.4612 [hep-ph]} \BibitemShut
  {NoStop}%
\bibitem [{\citenamefont {He}\ \emph {et~al.}(2011)\citenamefont {He},
  \citenamefont {Fries},\ and\ \citenamefont {Rapp}}]{He:2011yi}%
  \BibitemOpen
  \bibfield  {author} {\bibinfo {author} {\bibfnamefont {M.}~\bibnamefont
  {He}}, \bibinfo {author} {\bibfnamefont {R.~J.}\ \bibnamefont {Fries}}, \
  and\ \bibinfo {author} {\bibfnamefont {R.}~\bibnamefont {Rapp}},\ }\href
  {\doibase 10.1016/j.physletb.2011.06.019} {\bibfield  {journal} {\bibinfo
  {journal} {Phys. Lett. B}\ }\textbf {\bibinfo {volume} {701}},\ \bibinfo
  {pages} {445} (\bibinfo {year} {2011})},\ \Eprint
  {http://arxiv.org/abs/1103.6279} {arXiv:1103.6279 [nucl-th]} \BibitemShut
  {NoStop}%
\bibitem [{\citenamefont {Riek}\ and\ \citenamefont
  {Rapp}(2010)}]{Riek:2010fk}%
  \BibitemOpen
  \bibfield  {author} {\bibinfo {author} {\bibfnamefont {F.}~\bibnamefont
  {Riek}}\ and\ \bibinfo {author} {\bibfnamefont {R.}~\bibnamefont {Rapp}},\
  }\href {\doibase 10.1103/PhysRevC.82.035201} {\bibfield  {journal} {\bibinfo
  {journal} {Phys. Rev. C}\ }\textbf {\bibinfo {volume} {82}},\ \bibinfo
  {pages} {035201} (\bibinfo {year} {2010})},\ \Eprint
  {http://arxiv.org/abs/1005.0769} {arXiv:1005.0769 [hep-ph]} \BibitemShut
  {NoStop}%
\bibitem [{\citenamefont {He}\ \emph {et~al.}(2012)\citenamefont {He},
  \citenamefont {Fries},\ and\ \citenamefont {Rapp}}]{He:2011qa}%
  \BibitemOpen
  \bibfield  {author} {\bibinfo {author} {\bibfnamefont {M.}~\bibnamefont
  {He}}, \bibinfo {author} {\bibfnamefont {R.~J.}\ \bibnamefont {Fries}}, \
  and\ \bibinfo {author} {\bibfnamefont {R.}~\bibnamefont {Rapp}},\ }\href
  {\doibase 10.1103/PhysRevC.86.014903} {\bibfield  {journal} {\bibinfo
  {journal} {Phys. Rev. C}\ }\textbf {\bibinfo {volume} {86}},\ \bibinfo
  {pages} {014903} (\bibinfo {year} {2012})},\ \Eprint
  {http://arxiv.org/abs/1106.6006} {arXiv:1106.6006 [nucl-th]} \BibitemShut
  {NoStop}%
\bibitem [{\citenamefont {Workman}\ \emph {et~al.}(2022)\citenamefont {Workman}
  \emph {et~al.}}]{PDG:2022pth}%
  \BibitemOpen
  \bibfield  {author} {\bibinfo {author} {\bibfnamefont {R.~L.}\ \bibnamefont
  {Workman}} \emph {et~al.} (\bibinfo {collaboration} {Particle Data Group}),\
  }\href {\doibase 10.1093/ptep/ptac097} {\bibfield  {journal} {\bibinfo
  {journal} {PTEP}\ }\textbf {\bibinfo {volume} {2022}},\ \bibinfo {pages}
  {083C01} (\bibinfo {year} {2022})}\BibitemShut {NoStop}%
\bibitem [{\citenamefont {Adamczyk}\ \emph
  {et~al.}(2016{\natexlab{b}})\citenamefont {Adamczyk} \emph
  {et~al.}}]{Adamczyk:2015fum}%
  \BibitemOpen
  \bibfield  {author} {\bibinfo {author} {\bibfnamefont {L.}~\bibnamefont
  {Adamczyk}} \emph {et~al.} (\bibinfo {collaboration} {STAR Collaboration}),\
  }\href {\doibase 10.1103/PhysRevC.93.014907} {\bibfield  {journal} {\bibinfo
  {journal} {Phys. Rev. C}\ }\textbf {\bibinfo {volume} {93}},\ \bibinfo
  {pages} {014907} (\bibinfo {year} {2016}{\natexlab{b}})},\ \Eprint
  {http://arxiv.org/abs/1509.08397} {arXiv:1509.08397 [nucl-ex]} \BibitemShut
  {NoStop}%
\bibitem [{\citenamefont {Nayak}(2021)}]{Nayak:2020djj}%
  \BibitemOpen
  \bibfield  {author} {\bibinfo {author} {\bibfnamefont {K.}~\bibnamefont
  {Nayak}} (\bibinfo {collaboration} {STAR Collaboration}),\ }\href {\doibase
  10.1016/j.nuclphysa.2020.121855} {\bibfield  {journal} {\bibinfo  {journal}
  {Nucl. Phys. A}\ }\textbf {\bibinfo {volume} {1005}},\ \bibinfo {pages}
  {121855} (\bibinfo {year} {2021})},\ \Eprint
  {http://arxiv.org/abs/2002.12066} {arXiv:2002.12066 [nucl-ex]} \BibitemShut
  {NoStop}%
\bibitem [{\citenamefont {Adam}\ \emph
  {et~al.}(2020{\natexlab{b}})\citenamefont {Adam} \emph
  {et~al.}}]{STAR:2019ank}%
  \BibitemOpen
  \bibfield  {author} {\bibinfo {author} {\bibfnamefont {J.}~\bibnamefont
  {Adam}} \emph {et~al.} (\bibinfo {collaboration} {STAR Collaboration}),\
  }\href {\doibase 10.1103/PhysRevLett.124.172301} {\bibfield  {journal}
  {\bibinfo  {journal} {Phys. Rev. Lett.}\ }\textbf {\bibinfo {volume} {124}},\
  \bibinfo {pages} {172301} (\bibinfo {year} {2020}{\natexlab{b}})},\ \Eprint
  {http://arxiv.org/abs/1910.14628} {arXiv:1910.14628 [nucl-ex]} \BibitemShut
  {NoStop}%
\bibitem [{\citenamefont {Adam}\ \emph {et~al.}(2021)\citenamefont {Adam} \emph
  {et~al.}}]{STAR:2021tte}%
  \BibitemOpen
  \bibfield  {author} {\bibinfo {author} {\bibfnamefont {J.}~\bibnamefont
  {Adam}} \emph {et~al.} (\bibinfo {collaboration} {STAR Collaboration}),\
  }\href {\doibase 10.1103/PhysRevLett.127.092301} {\bibfield  {journal}
  {\bibinfo  {journal} {Phys. Rev. Lett.}\ }\textbf {\bibinfo {volume} {127}},\
  \bibinfo {pages} {092301} (\bibinfo {year} {2021})},\ \Eprint
  {http://arxiv.org/abs/2101.11793} {arXiv:2101.11793 [hep-ex]} \BibitemShut
  {NoStop}%
\bibitem [{\citenamefont {Adams}\ \emph {et~al.}(2004)\citenamefont {Adams}
  \emph {et~al.}}]{Adams:2003am}%
  \BibitemOpen
  \bibfield  {author} {\bibinfo {author} {\bibfnamefont {J.}~\bibnamefont
  {Adams}} \emph {et~al.} (\bibinfo {collaboration} {STAR Collaboration}),\
  }\href {\doibase 10.1103/PhysRevLett.92.052302} {\bibfield  {journal}
  {\bibinfo  {journal} {Phys. Rev. Lett.}\ }\textbf {\bibinfo {volume} {92}},\
  \bibinfo {pages} {052302} (\bibinfo {year} {2004})},\ \Eprint
  {http://arxiv.org/abs/nucl-ex/0306007} {arXiv:nucl-ex/0306007} \BibitemShut
  {NoStop}%
\bibitem [{\citenamefont {Abdallah}\ \emph {et~al.}(2021)\citenamefont
  {Abdallah} \emph {et~al.}}]{STAR:2021twy}%
  \BibitemOpen
  \bibfield  {author} {\bibinfo {author} {\bibfnamefont {M.}~\bibnamefont
  {Abdallah}} \emph {et~al.} (\bibinfo {collaboration} {STAR Collaboration}),\
  }\href {\doibase 10.1103/PhysRevC.103.064907} {\bibfield  {journal} {\bibinfo
   {journal} {Phys. Rev. C}\ }\textbf {\bibinfo {volume} {103}},\ \bibinfo
  {pages} {064907} (\bibinfo {year} {2021})},\ \Eprint
  {http://arxiv.org/abs/2103.09451} {arXiv:2103.09451 [nucl-ex]} \BibitemShut
  {NoStop}%
\bibitem [{\citenamefont {Abelev}\ \emph {et~al.}(2015)\citenamefont {Abelev}
  \emph {et~al.}}]{Abelev:2014pua}%
  \BibitemOpen
  \bibfield  {author} {\bibinfo {author} {\bibfnamefont {B.~B.}\ \bibnamefont
  {Abelev}} \emph {et~al.} (\bibinfo {collaboration} {ALICE Collaboration}),\
  }\href {\doibase 10.1007/JHEP06(2015)190} {\bibfield  {journal} {\bibinfo
  {journal} {JHEP}\ }\textbf {\bibinfo {volume} {06}},\ \bibinfo {pages} {190}
  (\bibinfo {year} {2015})},\ \Eprint {http://arxiv.org/abs/1405.4632}
  {arXiv:1405.4632 [nucl-ex]} \BibitemShut {NoStop}%
\bibitem [{\citenamefont {Abelev}\ \emph
  {et~al.}(2013{\natexlab{a}})\citenamefont {Abelev} \emph
  {et~al.}}]{Abelev:2013vea}%
  \BibitemOpen
  \bibfield  {author} {\bibinfo {author} {\bibfnamefont {B.}~\bibnamefont
  {Abelev}} \emph {et~al.} (\bibinfo {collaboration} {ALICE Collaboration}),\
  }\href {\doibase 10.1103/PhysRevC.88.044910} {\bibfield  {journal} {\bibinfo
  {journal} {Phys. Rev. C}\ }\textbf {\bibinfo {volume} {88}},\ \bibinfo
  {pages} {044910} (\bibinfo {year} {2013}{\natexlab{a}})},\ \Eprint
  {http://arxiv.org/abs/1303.0737} {arXiv:1303.0737 [hep-ex]} \BibitemShut
  {NoStop}%
\bibitem [{\citenamefont {Abelev}\ \emph
  {et~al.}(2013{\natexlab{b}})\citenamefont {Abelev} \emph
  {et~al.}}]{Abelev:2013qoq}%
  \BibitemOpen
  \bibfield  {author} {\bibinfo {author} {\bibfnamefont {B.}~\bibnamefont
  {Abelev}} \emph {et~al.} (\bibinfo {collaboration} {ALICE Collaboration}),\
  }\href {\doibase 10.1103/PhysRevC.88.044909} {\bibfield  {journal} {\bibinfo
  {journal} {Phys. Rev. C}\ }\textbf {\bibinfo {volume} {88}},\ \bibinfo
  {pages} {044909} (\bibinfo {year} {2013}{\natexlab{b}})},\ \Eprint
  {http://arxiv.org/abs/1301.4361} {arXiv:1301.4361 [nucl-ex]} \BibitemShut
  {NoStop}%
\bibitem [{\citenamefont {Moreau}\ \emph {et~al.}(2021)\citenamefont {Moreau},
  \citenamefont {Soloveva}, \citenamefont {Grishmanovskii}, \citenamefont
  {Voronyuk}, \citenamefont {Oliva}, \citenamefont {Song}, \citenamefont
  {Kireyeu}, \citenamefont {Coci},\ and\ \citenamefont
  {Bratkovskaya}}]{Moreau:2021clr}%
  \BibitemOpen
  \bibfield  {author} {\bibinfo {author} {\bibfnamefont {P.}~\bibnamefont
  {Moreau}}, \bibinfo {author} {\bibfnamefont {O.}~\bibnamefont {Soloveva}},
  \bibinfo {author} {\bibfnamefont {I.}~\bibnamefont {Grishmanovskii}},
  \bibinfo {author} {\bibfnamefont {V.}~\bibnamefont {Voronyuk}}, \bibinfo
  {author} {\bibfnamefont {L.}~\bibnamefont {Oliva}}, \bibinfo {author}
  {\bibfnamefont {T.}~\bibnamefont {Song}}, \bibinfo {author} {\bibfnamefont
  {V.}~\bibnamefont {Kireyeu}}, \bibinfo {author} {\bibfnamefont
  {G.}~\bibnamefont {Coci}}, \ and\ \bibinfo {author} {\bibfnamefont
  {E.}~\bibnamefont {Bratkovskaya}},\ }\href {\doibase 10.1002/asna.202113988}
  {\bibfield  {journal} {\bibinfo  {journal} {Astron. Nachr.}\ }\textbf
  {\bibinfo {volume} {342}},\ \bibinfo {pages} {715} (\bibinfo {year}
  {2021})},\ \Eprint {http://arxiv.org/abs/2101.05688} {arXiv:2101.05688
  [nucl-th]} \BibitemShut {NoStop}%
\end{thebibliography}%

\end{document}